\documentclass[a4paper,10pt,openany]{article}
\usepackage{pst-3d}
\usepackage{pst-3dplot}
\usepackage{etex}
\usepackage[utf8]{inputenc}
\usepackage[english]{babel}
\usepackage{amsmath,amssymb,amsthm,mathrsfs,amsfonts,dsfont}
\usepackage{graphicx}
\usepackage[small, skip=10pt]{caption}
\usepackage{epigraph}
\usepackage{indentfirst}
\usepackage{booktabs}
\usepackage{fancyhdr}
\usepackage{vmargin}
\usepackage{feynmf}
\usepackage{yfonts}
\usepackage{lscape}
\usepackage{subfig}
\usepackage{textcomp}
\usepackage{pstricks}
\usepackage[metapost]{mfpic}
\usepackage{fancybox}
\usepackage{slashed}
\usepackage[verbose]{wrapfig}
\usepackage{pifont}
\usepackage{keystroke}
\usepackage{appendix}
\usepackage{enumitem}
\usepackage{array}
\usepackage{colortbl}
\usepackage{alltt}
\usepackage{boxedminipage}
\usepackage{tikz}
\usepackage{calc}
\usepackage{subfloat}
\usepackage{multirow}
\usepackage{smartdiagram}
\usepackage{cmll}
\usepackage{multicol}
\definecolor{color1}{RGB}{204,0,51}
\definecolor{color2}{RGB}{159,182,205}
\usepackage{bookmark}

\usepackage{verbatim}
\usepackage[vcentermath]{youngtab}
\usepackage{young}
\usepackage{pst-grad} 
\usepackage{pst-plot}
\usepackage{transparent}
\usepackage{color,soul}
\usepackage{cancel}
\usepackage{bm}
\usepackage{pifont}
\usepackage{placeins}
\usepackage{mathtools}
\usepackage{rotating}
\usepackage[refpage]{nomencl}
\usepackage{hhline}
\usepackage{marginnote}
\usepackage{upgreek}
\usepackage{stackengine}
\usepackage{setspace}
\usepackage{ytableau}
\usepackage{listings}
\usepackage{calligra}
\usepackage{hyperref}
\usepackage{afterpage}
\usepackage{cite}
\usepackage{cleveref}
\usepackage{makecell}
\usepackage{accents}
\usepackage{breqn}
\usepackage{environ}

\usepackage{tikz-cd}
\usetikzlibrary{cd}

\usepackage[final]{showlabels}

\def\rme{\mathrm{e}}
\def\rmd{\mathrm{d}}
\def\rmR{\mathrm{R}}

\definecolor{darkergreen}{rgb}{0.0, 0.5, 0.0}

\allowdisplaybreaks

\sethlcolor{yellow}
\definecolor{boh}{RGB}{79,47,79}

\hypersetup{
linkbordercolor={red}
}

\hypersetup{
    bookmarks=false,         
    unicode=false,          
    pdftoolbar=true,        
    pdfmenubar=true,        
    pdffitwindow=false,     
    pdfstartview={FitH},    
    pdftitle={Non-Relativistic String Theory and Branes},    
    pdfauthor={Eric Bergshoeff},     
    pdfsubject={},   
    pdfnewwindow=true,      
    colorlinks=false,       
    linkcolor=red,          
    citecolor=red,        
    filecolor=red,      
    urlcolor=red           
    linkbordercolor={red},
    citebordercolor={red},
    urlbordercolor={red}
}

\makeatletter

\newcommand{\Rmnum}[1]{\expandafter\@slowromancap\romannumeral #1@}
\makeatother

\theoremstyle{definition}

\theoremstyle{remark}

\theoremstyle{proposition}

\usepackage{alphalph}

\makeatletter
\newalphalph{\aalphalph}[mult]{\alphalph@alph}{26}
\newcommand{\alphalphval}[1]{%
  \@ifundefined{c@#1}{
    \aalphalph{#1}
  }{%
    \aalphalph{\value{#1}}
  }
}
\makeatother

\AtBeginEnvironment{subequations}{%
  \let\alph\alphalphval%
}

\usepackage{color}

\def\chapterautorefname~#1\null{Chap.~(#1)\null}
\def\sectionautorefname~#1\null{Sec.~(#1)\null}
\def\subsectionautorefname~#1\null{sub--Sec.~(#1)\null}
\def\figureautorefname~#1\null{Fig.~(#1)\null}
\def\tableautorefname~#1\null{Tab.~(#1)\null}
\def\equationautorefname~#1\null{eq.~(#1)\null}

\def\equationautorefname~#1\null{eq.~(#1)\null}

\NewEnviron{multieq}[1][2]{

\begin{multicols}{#1}
\begin{subequations}
\setlength{\abovedisplayskip}{-15pt}
\allowdisplaybreaks
\begin{align}
\BODY
\end{align}
\end{subequations}
\end{multicols}
\setlength{\parindent}{0pt}
}

\DeclareMathAlphabet\mathbfcal{OMS}{cmsy}{b}{n}

\usepackage[yyyymmdd]{datetime}

\title{\bf  The Supersymmetric Neveu-Schwarz Branes \\ of Non-Relativistic String Theory}
\date{}

\begin{document}

\begin{flushright}
\small
April 8th, 2022\\
\normalsize
\end{flushright}
{\let\newpage\relax\maketitle}
\maketitle
\def\equationautorefname~#1\null{eq.~(#1)\null}
\def\tableautorefname~#1\null{tab.~(#1)\null}

\vspace{0.8cm}

\begin{center}

\renewcommand{\thefootnote}{\alph{footnote}}
{\sl\large E.~A.~Bergshoeff$^{~1}$}\footnote{Email: {\tt e.a.bergshoeff[at]rug.nl}},
{\sl\large J.~Lahnsteiner$^{~1}$}\footnote{Email: {\tt j.m.lahnsteiner[at]outlook.com}},
{\sl\large L.~Romano$^{~1,4}$}\footnote{Email: {\tt lucaromano2607[at]gmail.com}} and
{\sl\large J.~Rosseel$^{~2,3}$}\footnote{Email: {\tt jan.rosseel[at]univie.ac.at}},

\setcounter{footnote}{0}
\renewcommand{\thefootnote}{\arabic{footnote}}

\vspace{0.5cm}

${}^1${\it Van Swinderen Institute, University of Groningen\\
Nijenborgh 4, 9747 AG Groningen, The Netherlands}\\
\vskip .2truecm
${}^2${\it Faculty of Physics, University of Vienna,\\
Boltzmanngasse 5, 1090, Vienna, Austria }\\
\vskip .2truecm
${}^3${\it Erwin Schr\"odinger International Institute for Mathematics and Physics, \\
University of Vienna, Boltzmanngasse 9, 1090, Vienna, Austria} \\
\vskip .2truecm
${}^4${\it Departamento de Electromagnetismo y Electrónica,Universidad de Murcia,\\
Campus de Espinardo, 30100 Murcia, Spain}\\

\vspace{1.5cm}

\vskip -.3truecm

{\bf Abstract}
\end{center}
\begin{quotation}
  {\small
  \noindent We construct the basic Neveu-Schwarz (NS) brane solutions of non-relativistic string theory using longitudinal T-duality as a solution generating technique. Extending the NS background fields to a supergravity multiplet, we verify that all solutions we find are half-supersymmetric. The two perturbative solutions we find both have an interpretation as the background geometry outside a string-like object. Correspondingly, we refer to these non-Lorentzian backgrounds as winding string and unwound string solution. Whereas the winding string is part of the on-shell spectrum of non-relativistic string theory, the unwound string only makes sense off-shell where it mediates the instantaneous gravitational force. Seen from the nine-dimensional point of view, we find that the winding string solution is sourced by a non-relativistic massive particle and that the unwound string solution is sourced by a massless Galilean particle of zero colour and spin. We explain how these two string solutions fit into a discrete lightcone quantization of string theory. We shortly discuss the basic NS five-brane and Kaluza-Klein monopole solutions and show that they are both half-supersymmetric.
  }
  \end{quotation}
  \vskip .3truecm
  \centerline{\it Dedicated to Mees de Roo}

  \newpage

  \tableofcontents

  \section{Introduction}

\noindent Recently, much progress has been made in extending non-relativistic (NR) closed bosonic string theory in a flat background \cite{Gomis:2000bd,Danielsson:2000gi} and special curved backgrounds  \cite{Gomis:2005pg} to a general curved background. These results have been obtained either by taking a NR limit \cite{Bergshoeff:2018yvt,Bergshoeff:2019pij,Bergshoeff:2021bmc,Bidussi:2021ujm} or by applying a null reduction \cite{Bidussi:2021ujm,Harmark:2017rpg,Harmark:2018cdl,Harmark:2019upf}.  Open strings in a general curved background have been considered as well \cite{Gomis:2020fui,Gomis:2020izd}. Other recent work on non-relativistic  strings in a general curved background can be found in
  \cite{Kluson:2018egd,Kluson:2018vfd,Kluson:2019ifd,Roychowdhury:2019qmp,Hartong:2021ekg}.

These new results have, in particular, led to an identification of the target space geometry of the Neveu-Schwarz (NS) sector of NR string theory. This geometry is an extended version of the torsional Newton-Cartan geometry encountered, e.g., in Lifshitz holography \cite{Christensen:2013lma} in the sense that (i) it has two preferred directions longitudinal to the string instead of a single preferred time direction and (ii) it realizes not only a stringy version of Galilean symmetries but also an-isotropic dilatations. We refer to this geometry as torsional string Newton-Cartan geometry, see appendix \ref{sec:TSNC} and \cite{Bergshoeff:2021bmc, Bidussi:2021ujm, Bergshoeff:2021xxx}.  A priori, this geometry has a generic non-vanishing intrinsic torsion. Constraining this intrinsic torsion leads to different classes of geometries like in the case of torsional Newton-Cartan geometries  \cite{Figueroa-OFarrill:2020gpr}.

It turns out that extending  the bosonic non-relativistic string theory to a supersymmetric one requires the following constraint on the intrinsic torsion \cite{Bergshoeff:2021tfn}
  \begin{align}\label{constraint2}
  \tau_{[\mu}{}^- \partial_\nu \tau_{\rho]}{}^- = 0\,,
  \end{align}
where $\tau_\mu{}^-$ is one of the two longitudinal Vielbein fields that characterizes torsional string Newton-Cartan geometry, see appendix \ref{sec:TSNC} and \cite{Bergshoeff:2021bmc,Bergshoeff:2021tfn} for more details. The resulting geometry is the one that is relevant in this paper and will be referred to as {\sl self-dual dilatation invariant string Newton-Cartan} geometry \cite{Bergshoeff:2021tfn}---or, DSNC${}^-$ geometry for short. Non-relativistic strings then couple to the background fields of DSNC${}^-$ geometry including a Kalb-Ramond and dilaton field, and all these background fields must satisfy equations of motion that ensure (one loop) quantum Weyl invariance of the string sigma model \cite{Gomis:2019zyu,Yan:2019xsf,Gallegos:2019icg}. We will call the gravity theory describing the dynamics of the background fields `DSNC${}^-$ gravity'. This can be derived as a limit of NS gravity \cite{Bergshoeff:2021bmc} and extended to a supersymmetric theory as shown in \cite{Bergshoeff:2021tfn}.

Extended objects with $p$ spatial directions, i.e., $p$-branes, have played a crucial role in the understanding of non-perturbative properties of relativistic string theory, see e.g.~\cite{Polchinski:1998rq}. It is, therefore, natural to have a closer look at the branes of NR string theory\,\footnote{See\cite{Gomis:2000bd} for an early investigation of branes in NR string theory.} to learn more about its non-perturbative properties and internal consistency. In this paper, we will only consider the closed string sector. For NR open-string dynamics on D-branes see \cite{Gomis:2020fui, Gomis:2020izd,Yan:2021hte}, and references therein. The purpose of this paper is to study NS brane-like solutions in non-relativistic string theory from a target space point of view by looking for brane solutions of the background field equations of motion. The explicit form of the latter---that accords with DSNC${}^-$ geometry---has been derived in our recent work \cite{Bergshoeff:2021bmc}.

In order to investigate the NS brane solutions of non-relativistic string theory we use the intriguing fact that it is related to the discrete lightcone quantization of relativistic string theory. This relation is known as longitudinal T-duality and identifies the respective winding and momentum modes of both theories. It was first pointed out by \cite{Klebanov:2000pp,Danielsson:2000gi,Gomis:2000bd} in the flat space theory and later generalized to curved backgrounds in \cite{Bergshoeff:2018yvt}. Longitudinal T-duality relates non-relativistic string sigma models on a DSNC${}^-$ background to a relativistic Polyakov model on an NS background with a null isometry. This leaves an imprint on the effective target space description, namely a relation between DSNC${}^-$ gravity \cite{Bergshoeff:2021bmc} and a theory that will henceforth be referred to as NS${}_0$ gravity. The latter is defined as NS gravity in the presence of a lightlike isometry \cite{Goroff:1983hc, Duval:1984cj, Julia:1994bs, Bergshoeff:2017dqq}---supplemented with a non-trivial constraint that is T-dual to the constraint \eqref{constraint2} on the intrinsic torsion in DSNC${}^-$ gravity.

In this work, we will use the longitudinal T-duality relation as a solution-generating technique. That is, we consider the brane solutions of NS${}_0$ gravity and map them to solutions of DSNC${}^-$ gravity via the appropriate T-duality rules. It is important to keep in mind that not all solutions of NS gravity are solutions of NS${}_0$ gravity (and thus DSNC${}^-$ gravity) since they do not all satisfy the additional constraint that is T-dual to \eqref{constraint2}.  Schematically, we thus have
  \begin{align}\label{eq:solutioNS}
      &\mathrm{solutions~of~NS}_0\mathrm{~gravity}&&\subset && \mathrm{solutions~of~NS~gravity}\,.
  \end{align}
To illustrate this point, let us mention that the fundamental string solution of NS gravity has a lightlike isometry direction and solves the additional constraint that is T-dual to \eqref{constraint2}. The anti-fundamental string, however, does not obey this constraint and is thus not a solution of the NS${}_0$ theory. This is akin to the general intuition that anti-particles decouple in the non-relativistic limit.

In this paper, we will consider the supersymmetric extension of DSNC${}^-$ gravity and NS${}_0$ gravity to DSNC${}^-$ supergravity and $\mathcal N=(1,0)_0$ supergravity, respectively. In particular, we will use these extensions to investigate the supersymmetry preserved by the brane solutions. A useful fact in this context is that all solutions of $\mathcal N=(1,0)$ supergravity that preserve some amount of supersymmetry also have a null Killing vector \cite{Fontanella:2019avn}. This implies that there are a large number of candidate solutions of NS${}_0$ supergravity, including fundamental, solitonic, and intersecting branes, suitable for generating T-dual solutions of DSNC${}^-$ supergravity. Here, we will restrict ourselves to the simplest solutions of DSNC${}^-$ supergravity and show that they are all half-supersymmetric. We will derive their Killing spinors  by using the Killing spinor equations of DSNC${}^- $supergravity derived in our recent work \cite{Bergshoeff:2021tfn}

Longitudinal T-duality applied to the pp-wave and fundamental string solution of NS gravity leads to the two basic NS string solutions of DSNC${}^-$ gravity. Both solutions are found to be half-supersymmetric. We refer to these as the winding and unwound string solution, respectively. We argue that the winding string solution is sourced by a winding mode of non-relativistic string theory---hence the name. From a nine-dimensional point of view this takes the form of a massive Bargmann particle and the associated supergravity solution describes the gravitational field outside this point particle. Similarly, we argue that the unwound string describes the field outside the momentum mode which can be described as a massless Galilean particle of zero colour and spin  \cite{souriau1970structure,Batlle:2017cfa} from the nine-dimensional point of view. It is known \cite{Danielsson:2000gi,Danielsson:2000mu,Gomis:2000bd} that the unwound momentum modes do not appear as asymptotic states in non-relativistic string theory. It was shown by \cite{Danielsson:2000mu} that these modes appear as intermediate states in scattering amplitudes giving rise to the instantaneous Newtonian gravitational interaction.\footnote{For this reason these modes were referred to as Newtonian gravitons in \cite{Danielsson:2000mu}. In this paper, we will not use this nomenclature and instead refer to them as unwound string states.} For completeness, we also consider two solitonic five-brane solutions of DSNC${}^-$ gravity.

This paper is organized as follows. In section \ref{sec:TDuality}, we will give a  target space description of longitudinal T-duality suitable for using this duality as a solution generating transformation. We also discuss two other notions of T-duality that are characteristic of non-relativistic string theory: transversal and lightlike T-duality. In section \ref{sec:susyTduality}, we study the supersymmetric generalization of longitudinal T-duality. In particular, we extend the bosonic background fields to supergravity background fields. Furthermore, we show how the longitudinal T-dual of the constraint \eqref{constraint2} of DSNC${}^-$ supergravity leads to a constraint of $\mathcal N=(1,0)_0$ supergravity that is required by consistency with supersymmetry. In section \ref{sec:susyBranes}, we apply the solution generating technique described above to the four basic NS solutions of NS${}_0$ gravity: the fundamental string, the pp-wave, the NS 5-brane, and the Kaluza-Klein monopole. In section \ref{sec:sources} we study the respective source terms corresponding to the solutions derived in section \ref{sec:susyBranes}. The structure of these source terms motivates an interpretation of the respective background geometries. Finally, in section \ref{sec:conclusions} we will give our conclusions. Appendix \ref{sec:conventions} contains our conventions and appendix \ref{sec:TSNC} gives some background information on torsional string Newton-Cartan and DSNC${}^-$ geometry.

\section{A Target Space Approach to Non-Relativistic T-duality}\label{sec:TDuality}

\noindent Relativistic closed string T-duality states that two closed string theories, compactified on a spatial circle of radius $R$, resp. $\ell_s^2/R$, describe the same physics.\footnote{Here and in the following $\ell_s$ denotes the string length.} T-duality can, for instance, be seen in closed bosonic string theory by examining the mass-shell and level-matching conditions that determine the spectrum of states of a string in a flat target space-time with one spatial direction compactified on a circle of radius $R$:
  \begin{align}
      \label{eq:spectrumcompbos}
      E^2 = \vec{p}^{\, 2} + \frac{n^2}{R^2} + \frac{w^2 R^2}{\ell_s^4} + \frac{2}{\ell_s^2} (N_L + N_R - 2) \,, \qquad \qquad \qquad N_L - N_R = n w \,.
  \end{align}
Here, $n, w \in \mathbb{Z}$, $E$ denotes the energy of the string states, and $\vec{p}$ their momentum in the non-compact directions. The integers $N_L$ and $N_R$ are the left- and right-moving string oscillator contributions. The second and third terms on the right-hand side of the mass-shell condition correspond to the quantized momentum and winding of the states along the compact direction, respectively. T-duality, i.e., the indistinguishability of two closed bosonic string theories compactified on a circle of radius $R$, resp. $\ell_s^2/R$, is then apparent since the spectrum determined by \eqref{eq:spectrumcompbos} is invariant under the interchanges
  \begin{align}
      R \quad \longleftrightarrow \quad \frac{\ell_s^2}{R} \,, \qquad \qquad \qquad n \quad \longleftrightarrow \quad w \,.
  \end{align}
  The states with $(w=1,n=0)$ and $(n=1,w=0)$ can be called the `fundamental string' and `pp-wave' states since they can be interpreted as sources for the homonymous solutions of the effective low-energy gravity theory of closed bosonic string theory. For $N_L=1=N_R$ the mass-shell condition \eqref{eq:spectrumcompbos} can be recognized as the dispersion relation of a massive particle in 25 dimensions. The masses of pure winding and momentum modes are $M_W=wR/\ell_s^2$ and $M_N = n/R$, respectively. This can be made even more explicit by performing a double dimensional reduction of the Polyakov model---while fixing either the winding number or the internal momentum along the compact direction.

  The same T-duality can also be phrased at the level of the string sigma model, coupled to a set of relativistic background fields, consisting of a target space metric $G_{\mu\nu}$, Kalb-Ramond field $B_{\mu\nu}$ and dilaton $\Phi$. T-duality then states that, assuming that the target space geometry has a spatial Killing vector $k^\mu\partial_\mu$, a string cannot distinguish between two different sets $\{{G}_{\mu\nu}, {B}_{\mu\nu},{\Phi}\}$ and $\{\tilde{G}_{\mu\nu},\tilde{B}_{\mu\nu},\tilde{\Phi}\}$ of background fields, provided these are related via the following involutive T-duality rules:
  \begin{align}\label{eq:Buscher}
      &\tilde G_{yy} = G_{yy}^{-1}\,, && \rme^{-2\tilde \Phi} = G_{yy}\,\rme^{-2\Phi}\,, \notag\\
      & \tilde G_{yi} =  G_{yy}^{-1}\,B_{yi}\,, && \tilde G_{ij} = G_{ij}-G_{yy}^{-1}\big(G_{yi}G_{yj}-B_{yi}B_{yj}\big)\,,\notag\\
      & \tilde B_{yi} = G_{yy}^{-1}\,G_{yi}\,, && \tilde B_{ij} = B_{ij} + 2\,G_{yy}^{-1}\,B_{y[i}G_{j]y}\,.
  \end{align}

  Here, we have split the spacetime coordinates $x^\mu$ as $x^\mu=(y,x^i)$, where $y$ is adapted to the Killing vector $k^\mu$, i.e. $k^\mu \partial_\mu = \partial_y$. We refer to appendix \ref{sec:conventions} for the naming conventions that we adopt for the coordinates adapted to the various isometries considered in this paper. The rules \eqref{eq:Buscher} are known as the Buscher rules for relativistic T-duality. For $G_{\mu\nu}$ and $B_{\mu\nu}$, they have been derived a long time ago from the string sigma model point of view as a worldsheet duality transformation \cite{Buscher:1987sk}. The corresponding T-duality shift of the dilaton $\Phi$ can be derived using a path integral approach \cite{Buscher:1987qj}. Under Buscher's T-duality rules \eqref{eq:Buscher}, the fundamental string and pp-wave gravity solutions that are sourced by the $w=1$, $n=0$ and $n=1$, $w=0$ perturbative string states are mapped into each other.

  In the presence of a spatial isometry, Buscher's rules \eqref{eq:Buscher} correspond to a $\mathbb{Z}_2$ symmetry of NS gravity, the bosonic sector common to the low energy effective actions of all ten-dimensional superstring theories \cite{Bergshoeff:1994cb}. This is most easily seen from a nine-dimensional point of view, by using the following reduction Ansatz to compactify the fields $\{{G}_{\mu\nu}, {B}_{\mu\nu},{\Phi}\}$ of NS gravity along the spatial Killing vector $k^\mu$:
  \begin{eqnarray}\label{Ansatzrel}
  E_\mu{}{}^{\hat A} = \bordermatrix{
  &a&9\cr
  i&e_i{}^{a}&k \, a_i\cr
  y&0&k}\,,\hskip .5truecm B_{yi} = b_i\,,\hskip .5truecm B_{ij} = b_{ij} + a_{[i}b_{j]}\,,\hskip .5truecm \Phi = \phi + \tfrac{1}{2}\log k\,.
  \end{eqnarray}
  In this Ansatz, $E_\mu{}^{\hat{A}}$ denotes a Vielbein for $G_{\mu\nu}$, all fields are assumed to be $y$-independent and we have furthermore split the ten-dimensional flat index $\hat{A}$ as $\hat A = (a,9)$. Note that setting $E_y{}^{a}=0$ in \eqref{Ansatzrel}, corresponds to gauge fixing the local Lorentz transformations with infinitesimal parameters $\Lambda^{a9}$, so that the remaining ones are nine-dimensional. Considering a second NS gravity theory with fields $\{\tilde{G}_{\mu\nu},\tilde{B}_{\mu\nu},\tilde{\Phi}\}$ and proposing an analogous Ansatz (obtained by putting tildes on all fields in \eqref{Ansatzrel}) for it, one finds that, upon using these Ans\"atze in \eqref{eq:Buscher}, Buscher's rules \eqref{eq:Buscher} amount to interchanging the two vectors $a_i$, $b_i$ and inverting the Kaluza-Klein scalar $k$:
  \begin{align}
      &\tilde k = k^{-1}\,, && \tilde a_i = b_i\,, && \tilde b_i = a_i\,,
  \end{align}
  corresponding to a $\mathbb{Z}_2$ symmetry of the nine-dimensional theory. In the case of a compactification on a circle, the scalar modulus $k$ represents the radius $R/\ell_s$ of the circle. Since $\rme^{\Phi_0} = g_s$, with $\Phi_0$ the asymptotic value of the dilaton $\Phi$ and $g_s$ the string coupling constant, Buscher's rules for the Kaluza-Klein scalar and the dilaton then imply that T-duality exchanges the compactification radius $R$ and $g_s$ for dual versions $\tilde{R}$ and the string coupling constant $\tilde{g}_s$ as follows:
  \begin{align}
  &R\quad \stackrel{T}{\longleftrightarrow} \quad \tilde R = \ell_s^2/R\,, && g_s\quad\stackrel{T}{\longleftrightarrow}\quad \tilde g_s = g_s\,\ell_s/R\,.
  \end{align}
  Applying a similar reasoning to string theories with $\mathcal N=2$ supersymmetry, one sees that T-duality reflects the property that for a compactification of type IIA string theory on a circle of radius $R$, there exists a dual compactification of type IIB theory on a circle of dual radius $\tilde R = \ell_s^2/R$, such that the IIA and IIB background fields reduce to the same components of the $9D$ maximal supergravity multiplet.

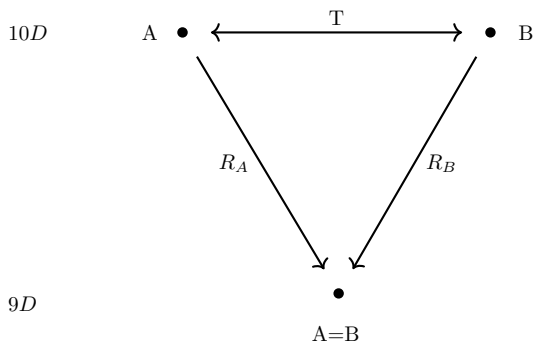
\begin{figure}[h]
   \centering
  \hskip 0truecm
  \begin{tikzpicture}[thick,scale=0.8, every node/.style={scale=0.8}]
      \node[draw,circle, fill=black,scale=0.2] (a) {A} ;
      \node[inner sep=0,minimum size=0,right of=a] (k) {};
      \node[draw,circle,right = 5 cm=k, fill=black,scale=0.2] (b) {B};
      \node[draw,circle,below =4.33cm,right=2.5 cm, fill=black, scale=0.2] (c) {C};
      \draw[<->,shorten <=9pt,shorten >=9pt] (a) edge node[above] {T} (b);
      \draw[->,shorten <=9pt,shorten >=9pt] (a) edge node[left] {\,$R_A$\,} (c);
      \draw[->,shorten <=9pt,shorten >=9pt]  (b) edge node[right] {\,$R_B$\,} (c);
      \filldraw (-0.8,-0.) node[anchor=west] {A};
      \filldraw (5.4,-0.) node[anchor=west] {B};
      \filldraw (2.,-5) node[anchor=west] {A=B};
      \filldraw (-3,-4.5) node[anchor=west] {9$D$};
      \filldraw (-3,0) node[anchor=west] {10$D$};
  \end{tikzpicture}
  \caption{T-duality provides a map between two supergravity theories A and B, each supplemented with a spatial isometry direction. The respective compact directions have a circumference $R_A=R$ and $R_B = \ell_s^2/R$, respectively. In this target space approach T-duality means that both theories A/B reduce to the same theory in nine dimensions. The T-duality rules are then effectively obtained by reducing theory A to nine dimensions and uplifting the result to theory B (or vice versa).}
  \label{fig:TDtriangle}
  \end{figure}

  Otherwise stated, $\mathcal N=2$ T-duality expresses the fact that IIA supergravity with a spatial isometry and IIB supergravity with a spatial isometry are both identical to $9D$ maximal supergravity.\,\footnote{The situation is slightly more subtle if the IIA Romans mass parameter is included. In that case, one should generalize the reduction of IIB supergravity to a so-called Scherk-Schwarz reduction  \cite{Bergshoeff:1996ui}. This subtlety will play no role in the present paper, where we only consider branes of low spatial extension.}
  Using this description of $\mathcal N=2$ T-duality provides a convenient way to derive the T-duality rules of all the RR fields \cite{Bergshoeff:1995as}. The general idea of this target space approach to T-duality is summarized in figure \ref{fig:TDtriangle}. For a microscopic discussion  of $\mathcal N=2$ T-duality, see \cite{Dai:1989ua,Dine:1989vu}.

It is natural to ask whether a similar notion of T-duality exists for non-relativistic string theory. Leaving aside the option of considering timelike T-duality \cite{Hull:1998vg}, there are three inequivalent notions of T-duality for non-relativistic string theory, distinguished by the nature of the Killing vector, along which the duality is performed. Specifically, this Killing vector can be tangent to a spatial longitudinal direction, a spatial transverse direction or a longitudinal lightlike direction. We will refer to and denote the three resulting T-dualities as `longitudinal', `transverse', and `lightlike', respectively. All of these will be defined more precisely below. Schematically, these three T-dualities relate the DSNC$^-$ and/or NS$_0$ backgrounds as follows:
\begin{subequations}\label{eq:TDualities}
      \begin{align}
          &\mathrm{longitudinal~T-duality}:&&\mathrm{DSNC}^-\qquad \longleftrightarrow\qquad \mathrm{NS}_0\,, \label{eq:TDualitiesL}\\
          &\mathrm{transversal~T-duality}:&&\mathrm{DSNC}^-\qquad \longleftrightarrow\qquad \mathrm{DSNC}^-\,, \label{eq:TDualitiesT}\\
          &\mathrm{lightlike~T-duality}:&&\mathrm{DSNC}_0^-\qquad \longleftrightarrow\qquad \mathrm{DSNC}_0^-\,. \label{eq:TDualitiesLL}
      \end{align}
\end{subequations}
Among these three we are most interested in the longitudinal T-duality, relating two seemingly different string theories: non-relativistic string theory on a DSNC${}^-$ background on the one side and relativistic string theory on a background with a null isometry on the other side. This was first pointed out in early works \cite{Klebanov:2000pp,Danielsson:2000gi,Gomis:2000bd} in flat space, relating non-relativistic string theory to the discrete lightcone quantization of string theory. Only recently has this relation been generalized to curved backgrounds \cite{Bergshoeff:2018yvt,Bergshoeff:2019pij}. Here, we will revisit the T-duality rules from a pure target space point of view. We will define all three T-dualities, by reducing both sides of the schemes \eqref{eq:TDualities} over particular isometry directions and requiring that the results match in nine dimensions.

Before treating each of the dualities of \eqref{eq:TDualities} in turn, let us first summarize the field content of the DSNC${}^-$ and NS${}_0$ theories that are involved. The fields of the DSNC${}^-$ theory describe a particular type of ten-dimensional non-Lorentzian geometry that is equipped with a distribution of rank 8. This distribution is defined by the existence of a doublet of one-forms $\tau_\mu{}^A$, with $A=0,1$ or $A=+,-$ (in case we use lightcone indices). The $\tau_\mu{}^A$ are referred to as longitudinal Vielbeine. On top of these, the geometry is given in terms of transverse spatial Vielbeine $e_\mu{}^{A'}$, with $A'=2,\cdots,9$, a Kalb-Ramond two-form $b_{\mu\nu}$ and a dilaton $\phi$. The fields of the NS${}_0$ theory are given by the usual Neveu-Schwarz fields, i.e. a metric $\tilde G_{\mu\nu}$, a Kalb-Ramond two-form $\tilde B_{\mu\nu}$ and a dilaton $\tilde\Phi$.

  In the following, we will provide details on how the T-dualities \eqref{eq:TDualities} can be defined from the viewpoint of particular dimensional reductions of the DSNC${}^-$ and/or NS${}_0$ theories involved.

  \subsection{Longitudinal T-duality}\label{sec:longitudinalTD}

  \noindent Non-relativistic longitudinal T-duality, see eq.~(\ref{eq:TDualitiesL}\hskip -.13truecm), refers to the fact that a non-relativistic string in a DSNC${}^-$ background, with an isometry in a spatial longitudinal worldsheet direction, is T-dual to a string in an NS$_0$ background, i.e.~an NS background with a lightlike isometry. Here, we will revisit the Buscher rules for longitudinal T-duality \cite{Bergshoeff:2018yvt,Bergshoeff:2019pij} from a target space point of view, by comparing the local symmetry transformation rules of the various nine-dimensional fields that arise upon reduction of DSNC${}^-$, respectively NS$_0$ gravity, along a longitudinal spatial, respectively lightlike isometry direction.

  Let us first recall that the fields of DSNC${}^-$ gravity transform under local longitudinal $\mathsf{SO}(1,1)$ transformations, transverse spatial $\mathsf{SO}(8)$ rotations, Galilean boosts, dilatations and a one-form gauge symmetry, with respective parameters $\lambda_M\,, \lambda^{A'B'}\,,\lambda^{AA'}\,, \lambda_D$ and $\theta_\mu$, as follows \cite{Bergshoeff:2021bmc}:
  \begin{align} \label{eq:symmsDSNCm}
   &\delta \tau_\mu{}^A =  \lambda_M\,\epsilon^A{}_B\tau_\mu{}^B + \Lambda_D\tau_\mu{}^A\,, && \delta e_\mu{}^{A'} =  \lambda^{A'}{}_{B'}e_\mu{}^{B'} - \lambda_A{}^{A'} \tau_\mu{}^A\,, \notag\\
   &\delta \phi = \lambda_D\,, && \delta b_{\mu\nu} =  2\partial_{[\mu}\theta_{\nu]}-2\,\epsilon_{AB}\lambda^A{}_{A'}\tau_{[\mu}{}^B e_{\nu]}{}^{A'}\,.
  \end{align}
  Furthermore, under diffeomorphisms (with parameters $\xi^\mu$), $\phi$ transforms as a scalar, $\tau_\mu{}^{A}$ and $e_\mu{}^{A'}$ as one-forms and $b_{\mu\nu}$ as a two-form.

  To consider longitudinal T-duality for a non-relativistic string in a DSNC${}^-$ geometry, we assume that the latter has a spatial isometry in a direction longitudinal to the string, whose Killing vector we denote by $k^\mu$. In adapted ten-dimensional coordinates $x^\mu = (z,x^i) = (z,t,z_{(8)}^m)$, in which $k^\mu\partial_\mu = \partial_z$, the condition that the isometry is longitudinal spatial can then be defined as
  \begin{align}
      &\tau_z{}^0 \equiv k^\mu\tau_\mu{}^0 = 0\,, && \tau_z{}^1 \equiv k^\mu\tau_\mu{}^1 = 1\,, && e_z{}^{A'} \equiv k^\mu e_\mu{}^{A'} =0\,.
  \end{align}
  Note that this fixes half of the Galilean boosts (namely those with parameters $\lambda^{1A'}$), the longitudinal $\mathsf{SO}(1,1)$ transformations, as well as the dilatations. We denote the remaining, unfixed field components of $\tau_\mu{}^A$, $e_\mu{}^{A'}$ and $b_{\mu\nu}$ as follows
  \begin{equation}
    \label{eq:longvielbeinansatz}
    \tau_\mu{}^A = \bordermatrix{ & 0 & 1 \cr
    i  &  \tau_i & - n_i \cr
    z  & 0 & 1}\,, \hskip .4truecm e_\mu{}^{A'} = \bordermatrix{& A'  \cr
  i & e_i{}^{A'}  \cr
  z & 0 }\,,\hskip .4truecm
    b_{zi} = m_i\,,\hskip .4truecm   b_{ij} = m_{ij} - 2\, n_{[i} m_{j]}\,.
  \end{equation}
  Assuming that the fields $\tau_i$, $n_i$, $e_i{}^{A'}$, $m_i$, $m_{ij}$, as well as $\phi$ are all $z$-independent, we can view \eqref{eq:longvielbeinansatz} as a reduction Ansatz and consider the symmetry transformation rules \eqref{eq:symmsDSNCm} from the nine-dimensional point of view. We decompose and rename the ($z$-independent) unfixed parameters of diffeomorphisms, one-form symmetries and Galilean boosts as
  \begin{align} \label{eq:longparameters}
    &\xi^\mu = \{\xi^z= - \alpha,\xi^i\}\,, && \theta_\mu = \{\theta_z=-\beta, \theta_i\}\,, && \lambda_{A'} = -\lambda_{0 A'}\,.
  \end{align}
  Using \eqref{eq:longvielbeinansatz}, \eqref{eq:longparameters} in \eqref{eq:symmsDSNCm}, we then find that the nine-dimensional fields transform as follows under the unfixed symmetries:
  \begin{align} \label{eq:symmetriesiso}
        &\delta\tau_i=0\,,&&\delta\phi = 0\,, \nonumber \\
        & \delta m_i=\partial_i\beta -\lambda_{A'}e_{i}{}^{A'} \,,&& \delta n_i = \partial_i\alpha\,, \nonumber \\
        &\delta e_{i}{}^{A'} = \lambda^{A'}{}_{B'}e_{i}{}^{B'}+\lambda^{A'}\tau_{i}\,, && \delta m_{ij} = 2\,\partial_{[i}\theta_{j]} + 2\,n_{[i} \partial_{j]} \beta \,,
  \end{align}
  where we have not indicated the lower-dimensional general coordinate transformations with parameters $\xi^i$, since they act in the usual way. From these transformation rules, one sees that the longitudinal spatial reduction of DSNC${}^-$ gravity leads to two sets of fields $(\tau_i, e_{i}{}^{A'}, m_i)$ and $(\phi, n_i, m_{ij})$ that both realize the nine-dimensional Bargmann algebra, i.e. the centrally extended Galilei algebra (with central extension parameter $\beta$). The fields $(\tau_i, e_{i}{}^{A'}, m_i)$ have the transformation rules of the geometric fields that define a nine-dimensional torsional Newton-Cartan geometry. The remaining fields $(\phi,n_i,m_{ij})$, where $m_{ij}$ transforms non-trivially under the central extension of the Bargmann algebra, constitute a matter multiplet.

  We now consider the T-dual theory given by NS${}_0$ gravity. The sublabel $``0"$ indicates that the theory has a lightlike isometry, so that it is effectively nine-dimensional. Furthermore, one also needs to impose a bosonic constraint that is T-dual to eq. \eqref{constraint2}. We will defer a discussion of this constraint until the end of this section; for now, let us focus on how the presence of a lightlike Killing vector $k^\mu$ changes the local symmetry structure of the theory. To see this, we recall the transformation rules of relativistic NS gravity under Lorentz transformations, with parameters $\tilde\Lambda^{\hat A}{}_{\hat B}$, and one-form gauge transformations, with parameters $\tilde\Theta_\mu$:
  \begin{align} \label{eq:NStrafos}
  &\delta \tilde E_\mu{}^{\hat A} =  \tilde \Lambda^{\hat A}{}_{\hat B}\tilde E_\mu{}^{\hat B}\,,&& \delta \tilde B_{\mu\nu} =  2\partial_{[\mu}\tilde\Theta_{\nu]}\,,&& \delta \tilde \Phi =0\,.
  \end{align}
  The fields $\tilde E_\mu{}^{\hat A}$, $\tilde B_{\mu\nu}$ and $\tilde{\Phi}$ moreover transform under diffeomorphisms, with parameters $\tilde{\xi}^\mu$, as one-forms, a two-form and a scalar, respectively.
  We will choose adapted coordinates $x^\mu=(z,x^i)$, such that $k^\mu \partial_\mu = \partial_z$. Setting $\tilde{E}_z{}^{A'} = 0$ to gauge fix some of the local Lorentz transformations, the lightlike nature of $k^\mu$ then implies that $k^\mu k_\mu = \tilde G_{zz} = -2\,\tilde E_z{}^+\tilde E_z{}^- = 0$, which is without loss of generality solved by $\tilde E_z{}^- = 0$. The condition $\tilde{E}_z{}^{A'} = 0$ is then seen to be fixing the Lorentz transformations with parameters $\tilde\Lambda^{-A'}$. We can also fix the boosts with parameters $\tilde\Lambda^{+-}$, by setting $\tilde E_z{}^+ = 1$. Taking all this into account, we parametrize the remaining field components via the following convenient Ansatz \cite{Julia:1994bs,Bergshoeff:2017dqq}:
  \begin{align}
    \label{eq:vielbeinansatz}
   \tilde E_\mu{}^{\hat A} =
   \bordermatrix{& A' & - & + \cr
  i & \tilde e_i{}^{A'} &  \tilde\tau_i & - \tilde m_i \cr
  z & 0 & 0 & 1}\,, \hskip .4truecm  \tilde B_{zi} = \tilde n_i\,,\hskip .4truecm   \tilde B_{ij} = \tilde m_{ij} \,,\hskip .4truecm \tilde \Phi = \tilde \phi\,,
  \end{align}
  where the nine-dimensional fields $\tilde e_i{}^{A'}$, $\tilde \tau_i$, $\tilde m_i$, $\tilde n_i$, $\tilde m_{ij}$ and $\tilde\phi$ are all $z$-independent.
  In order to analyze the local symmetries of the nine-dimensional theory, we decompose and rename the remaining ($z$-independent) relevant symmetry parameters as follows:
  \begin{align} \label{eq:parNS0}
    &\tilde\xi^\mu = \{\tilde\xi^z=-\tilde\beta,\tilde\xi^i\}\,,&& \tilde\Theta_\mu = \{\tilde\Theta_z=-\tilde\alpha, \tilde\theta_i\}\,,&&  \tilde{\lambda}_{A'} = \tilde{\Lambda}_{A'-}\,,&&\tilde{\lambda}_{A'B'}=\tilde{\Lambda}_{A'B'}\,.
  \end{align}
  Upon using \eqref{eq:vielbeinansatz} and \eqref{eq:parNS0}, we then find that the rules \eqref{eq:NStrafos} give rise to the following nine-dimensional transformation rules:
  \begin{align}\label{eq:nullsymmetriesiso}
        &\delta\tilde\tau_i=0\,,&&\delta\tilde\phi = 0\,, \nonumber \\
        &\delta \tilde m_i = \partial_i\tilde\beta - \tilde\lambda_{A'}\tilde{e}_{i}{}^{A'} \,,&&\delta \tilde n_i=\partial_i\tilde\alpha\,, \nonumber \\
          &\delta\tilde e_{i}{}^{A'} = \tilde{\lambda}^{A'}\tilde{\tau}_{i}+\tilde{\lambda}^{A'}{}_{B'}\tilde{e}_{i}{}^{B'}\,, && \delta\tilde m_{ij} = 2\,\partial_{[i}\tilde\theta_{j]} + 2\,\tilde{n}_{[i} \partial_{j]} \tilde{\beta} \,,
  \end{align}
  where we have not written the transformation rules under nine-dimensional diffeomorphisms (with parameters $\tilde{\xi}^i$), since they correspond to the usual ones. Like the longitudinal spatial reduction of DSNC${}^-$ gravity, the lightlike reduction of NS$_0$ gravity thus leads to two sets of fields $(\tilde\tau_i,\tilde e_{i}{}^{A'},\tilde m_i)$ and $(\tilde{\phi}, \tilde{n}_i, \tilde{m}_{ij})$, corresponding to the geometric fields of a nine-dimensional torsional Newton-Cartan geometry and the fields of a matter multiplet respectively.

  Longitudinal T-duality can now be interpreted as the statement that the longitudinal spatial reduction of a DSNC${}^-$ geometry and the lightlike reduction of an NS${}_0$ background both give rise to the same matter coupled torsional Newton-Cartan geometry in nine dimensions. The longitudinal T-duality map \eqref{eq:TDualitiesL} between DSNC$^-$ and NS$_0$ gravity can then be specified by the following rules that map the transformation rules \eqref{eq:symmetriesiso} and \eqref{eq:nullsymmetriesiso} to each other
  \begin{align}\label{identifications}
  &\tilde \tau_i = \tau_i\,, && \tilde\phi = \phi\,,&&\tilde e_{i}{}^{A'} = e_{i}{}^{A'},\notag\\
  &\tilde m_i = m_i\,, && \tilde n_i = n_i\,, && \tilde m_{ij} =  m_{ij}\,,
  \end{align}
  along with the identifications $\tilde\alpha = \alpha$, $\tilde\beta = \beta$, $\tilde\lambda_{A'} = \lambda_{A'}$ of the symmetry parameters. The non-relativistic T-duality rules are then obtained by uplifting the identification in nine dimensions to ten dimensions. Here we choose to express these as a map from the DSNC$^-$ to the NS${}_0$ theory:
    \begin{align}\label{eq:BuscherRescal}
      &\tilde G_{zz} = 0\,, && \tilde\Phi = \phi\,, \nonumber \\
      &\tilde E_i{}^{A'} = e_i{}^{A'}\,,&& \tilde E_i{}^-= \tau_i{}^0\,, && \tilde E_i{}^+ = -b_{zi} \,,\nonumber \\
      &\tilde B_{zi} = -\tau_i{}^1\,, && \tilde B_{ij} = b_{ij} + 2\,\mathcal\,b_{z[i}\tau_{j]}{}^1\,.
    \end{align}
  It is straightforward to invert these rules and establish Buscher rules that relate NS${}_0$ to DSNC${}^-$ backgrounds. In particular,
  \begin{align}
  b_{ij} =  \tilde B_{ij} + 2\,\tilde B_{z[i}\tilde E_{j]}{}^+\,,
  \end{align}
  while the inverses of the other rules are trivial. Note that we have fixed the local dilatation symmetry present in the DSNC${}^-$ theory by setting $\tau_z{}^1=1$. Loosely speaking, this is T-dual to the vanishing of the would-be Kaluza-Klein scalar $\tilde G_{zz}$. In that sense, we can say that DSNC${}^-$ gravity with a longitudinal spatial isometry is identical to NS gravity with a null isometry. We also remark that in this case of longitudinal T-duality, one cannot associate an independent modulus that represents the radius of a compactification circle. At the DSNC${}^-$ gravity side, this is due to the gauge-fixing $\tau_z{}^1=1$, while at the NS${}_0$ side, it is due to the fact that the radius of a circle in a null direction can be arbitrarily changed by a local Galilean boost transformation, see \cite{Seiberg:1997ad}. In fact, the invariant length of the lightlike circle is zero since $\tilde G_{zz}=0$.  The only independent modulus is the string coupling constant $g_s$ contained in the dilaton field, which is invariant under T-duality.

  \begin{figure}[t]
   \centering
  \hskip 0truecm
  \begin{tikzpicture}[thick,scale=0.8, every node/.style={scale=0.8}]
      \node[draw,circle, fill=black,scale=0.2] (a) {A} ;
      \node[inner sep=0,minimum size=0,right of=a] (k) {};
      \node[draw,circle,right = 5 cm=k, fill=black,scale=0.2] (b) {B};
      \node[draw,circle,below =4.33cm,right=2.5 cm, fill=black, scale=0.2] (c) {C};
      \draw[<->,shorten <=9pt,shorten >=9pt] (a) edge node[above] {longitudinal T-duality} (b);
      \draw[->,shorten <=9pt,shorten >=9pt] (a) edge node[left] {\,} (c);
      \draw[->,shorten <=9pt,shorten >=9pt]  (b) edge node[right] {\,} (c);
      \filldraw (-1.2, 1.) node[anchor=east] {\bfseries DSNC${}^-$};
      \filldraw (-0.5,-0.) node[anchor=east] {$(\tau_{\mu}{}^{A},\,e_{\mu}{}^{A'},\,b_{\mu\nu},\phi)$};
      \filldraw (-0.7,-0.9) node[anchor=east] {$\tau_{[\mu}{}^-\partial_\nu^{}\tau_{\rho]}{}^-=0$};
      \filldraw (7.2, 1.) node[anchor=east] {\bfseries NS${}_0$};
      \filldraw (5.5,-0.) node[anchor=west] {$(\tilde{E}_{\mu}{}^{\hat{A}},\,\tilde{B}_{\mu\nu},\,\tilde{\Phi})$};
      \filldraw (5.8,-0.9) node[anchor=west] {$\partial_{[\mu}\tilde Z_{\nu]}=0$};
      \filldraw (2.6,-5) node[anchor=center] {$(\tau_i,\, e_{i}{}^{A'},\, m_i, \phi,\, n_i,\, m_{ij})\quad \longleftrightarrow\quad(\tilde\tau_i,\,\tilde e_{i}{}^{A'},\,\tilde m_i,\, \tilde{\phi},\, \tilde{n}_i,\, \tilde{m}_{ij})$};
      \filldraw (2.6,-5.9) node[anchor=center] {$\partial_{[i}\tau_{j]} = -\partial_{[i}n_{j]}\quad \longleftrightarrow\quad\partial_{[i}\tilde\tau_{j]} = -\partial_{[i}\tilde n_{j]}$};
      \filldraw (5., -6.8) node[anchor=east] {\bfseries Torsional Newton-Cartan};
      \filldraw (-6,-4.5) node[anchor=west] {9$D$};
      \filldraw (-6,0) node[anchor=west] {10$D$};
  \end{tikzpicture}
  \caption{Schematic overview of the target space approach to longitudinal T-duality $T_L$. The non-relativistic T-duality rules \eqref{eq:BuscherRescal} are obtained by matching the nine-dimensional fields following from a null reduction of NS${}_0$ gravity with those following from a longitudinal reduction of DSNC${}^-$ gravity. Furthermore, we have indicated the respective geometric constraints: \eqref{constraint2} in the DSNC${}^-$ theory, \eqref{eq:NS0constraint} in the NS${}_0$ theory, and the nine-dimensional versions thereof.}
  \label{fig:TDtriangleFieldContent}
  \end{figure}
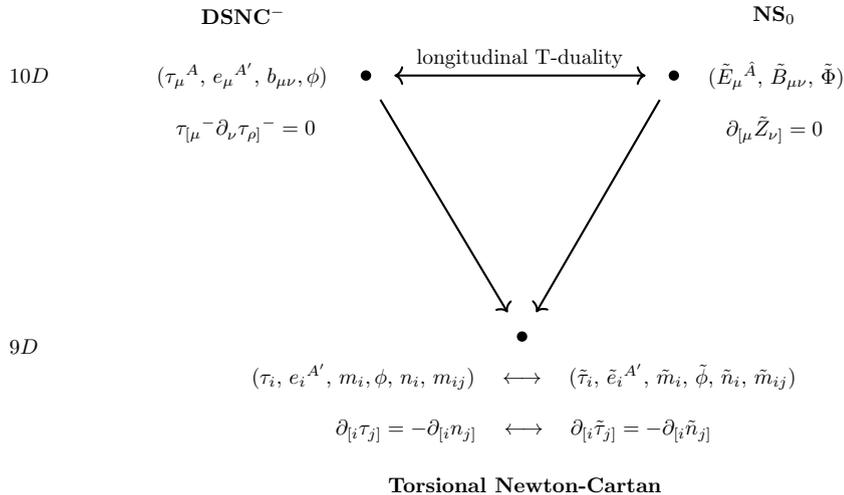

  In the introduction we have already mentioned that supersymmetry requires the geometric constraint eq.~\eqref{constraint2} for consistency \cite{Bergshoeff:2021tfn}. Applying the nine-dimensional identifications \eqref{identifications}, we can write this constraint as
  \begin{align}
      &\mathrm{longitudinal~T-duality}: && \tau^-\wedge\rmd\tau^- = 0\qquad\longleftrightarrow\qquad \tilde Z_{ij}= \partial_{[i}\tilde Z_{j]}=0\,,
  \end{align}
  where $\tilde Z_i = \tilde \tau_i + \tilde n_i$. Using that  $\tilde Z_z= \tilde{G}_{zz} =0$, this nine-dimensional constraint can be oxidized to NS${}_0$ gravity as follows:
  \begin{align}\label{eq:NS0constraint}
    &\partial_{[\mu}\tilde{Z}_{\nu]}=0\,,&&\mathrm{where} && \tilde{Z}_\mu = k^\nu\big(\tilde{G}_{\nu\mu}-\tilde{B}_{\nu\mu}\big)\,.
  \end{align}
  A schematic overview of the fields of the different nine-dimensional and ten-dimensional gravity theories, their corresponding constraints, and their relations under longitudinal T-duality can be found in Figure \ref{fig:TDtriangleFieldContent}.

  Geometrically, the condition \eqref{eq:NS0constraint} can be interpreted as an obstruction to the Killing vector being covariantly constant:
  \begin{align}
    \nabla_\mu k_\nu = - k^\rho \tilde{\mathcal{H}}_{\rho\mu\nu} \,,
  \end{align}
  where $\nabla_\mu$ is covariant with respect to the Levi-Civita connection and $\tilde{\mathcal{H}}=\rmd \tilde{B}$ denotes the Kalb-Ramond field strength.
  From the nine-dimensional point of view, the constraint $\partial_{[i}\tilde Z_{j]}=0$ expresses the fact that the time-like torsion $\partial_{[i}{\tilde \tau}_{j]}$ of the torsional Newton-Cartan geometry is furnished by the field strength of the $\mathsf{U}(1)$ gauge field $\tilde{n}_i$. This constraint has no obvious motivation in the purely bosonic target space theory. In section \ref{sec:susyTduality}, we will show that it follows from supersymmetry considerations when considering lightlike isometries in supersymmetric NS gravity. The inclusion of this additional condition will have important consequences when studying the solutions of the T-dual theories.

  To close this section, let us analyze the situation where the null Killing vector $k^\mu\partial_\mu$ is also covariantly constant:
  \begin{align} \label{eq:covconstk}
      \nabla_\mu k_\nu = 0\,.
  \end{align}
  Examples of NS backgrounds that exhibit such a null Killing vector are given by pp-wave solutions \cite{Brinkmann1}. In the adapted basis \eqref{eq:vielbeinansatz}, one has $k_\mu = \tilde G_{\mu\nu}k^\nu = \{k_z = 0,k_i = -\tilde{\tau}_i\}$. The antisymmetric part of \eqref{eq:covconstk} then implies that the nine-dimensional torsional Newton-Cartan geometry obeys the zero torsion constraint $\partial_{[i}\tilde{\tau}_{j]}=0$. The pp-wave geometries of NS gravity can thus be seen as matter-coupled torsionless Newton-Cartan geometries in one dimension lower. Note that the constraint $\partial_{[i} \tilde{Z}_{j]} = 0$ then also implies that the dimensional reduction of pp-waves leads to objects, for which the field strength of the $\mathsf{U}(1)$ gauge field $\tilde{n}_i$ is zero.

  \subsection{Transverse T-duality}\label{sec:transversalTD}

  \noindent We now consider the transverse T-duality \eqref{eq:TDualitiesT} that relates two DSNC${}^-$ theories, that are both T-dualized along an isometry direction that is transverse to the NR string. We work in adapted coordinates $x^\mu = (y,x^i)$, in which the Killing vector for the isometry reads $k^\mu\partial_\mu = \partial_y$. Transversality of the isometry means that
  \begin{align}
      & \tau_y{}^A \equiv k^\mu\tau_\mu{}^{A} = 0\,, && e_y{}^a \equiv k^\mu e_\mu{}^a = 0\,, && e_y{}^9 \equiv k^\mu e_\mu{}^9 = k \neq 0\,,
  \end{align}
  where we have split the eight-dimensional flat index $A^\prime$ as $A^\prime = (a,9)$ (without loss of generality).
  Compactifying in the spatial transverse $y$-direction, we make the following Kaluza-Klein Ansatz for the different DSNC${}^-$ gravity fields:
  \begin{align}\label{eq:transvielbeinansatz}
  &\tau_\mu{}^A = \bordermatrix{ & A  \cr
  i  &  \tau_i{}^A  \cr
  y  & 0 }\,, && e_\mu{}^{A'} = \bordermatrix{& a &9 \cr
  i & e_i{}^{a} &k\, n_i \cr
  y & 0& k }\,,\notag\\
  &b_{yi} = m_i\,, &&  b_{ij} = m_{ij} - m_{[i}n_{j]}\,,&&\rme^{2\phi} = k\,\rme^{2\varphi}\,.
  \end{align}
  It is not hard to see that the fields $(\tau_i{}^A, e_i{}^{A'}, m_{ij}, \varphi)$ realize a nine-dimensional DSNC${}^-$ geometry coupled to two $\mathsf{U}(1)-$vectors and a scalar $(m_i,n_i,k)$. We then denote the DSNC${}^-$ gravity fields of the transverse T-dual theory with a tilde and make an analogous Ansatz (with tildes) for these fields. Substituting these Ans\"atze into the transverse T-duality rules that were constructed in \cite{Bergshoeff:2018yvt,Bergshoeff:2019pij}, using a sigma model description of NR string theory, we find that the transverse T-duality amounts to a $\mathbb{Z}_2$ map from DSNC${}^-$ gravity onto itself, given by
  \begin{align}\label{transverseT}
        &\tilde k = k^{-1}\,, && \tilde m_i = n_i\,, && \tilde n_i = m_i\,.
  \end{align}
  Transverse T-duality thus acts by inverting the scalar $k$ and by interchanging the one-form fields $m_i$ and $n_i$.

  \subsection{Lightlike T-duality}\label{sec:lightlikeTD}

  \noindent Let us finally consider the lightlike T-duality (\ref{eq:TDualitiesLL}\hskip -.13truecm) between two DSNC$^-$ geometries, that is for both backgrounds performed along an isometry, whose Killing vector is lightlike in the directions longitudinal to the NR string. We choose adapted coordinates $x^\mu = (u, x^i)= (u,v,x^m)$ so that the Killing vector can be written as $\ell^\mu\partial_\mu = \partial_u$. Without loss of generality we choose
  \begin{align} \label{eq:DSNClightlike}
      &\tau_u{}^+ \equiv \ell^\mu\tau_\mu{}^+ = \ell \neq 0\,,&& \tau_u{}^- \equiv \ell^\mu\tau_\mu{}^- = 0\,, && e_u{}^{A'} \equiv \ell^\mu e_\mu{}^{A'}=0\,.
  \end{align}
  Note that this choice fixes the local boost symmetries with parameters $\lambda^{- A'}$.
  In the following we will refer to $\ell=\tau_u{}^+$ as the Kaluza-Klein scalar. Note however that it is not invariant under dilatations and $\mathsf{SO}(1,1)$ transformations and could thus be gauge fixed to $\ell=1$. We then propose the following Ansatz for the remaining field components of a DSNC$^-$ background:
  \begin{align} \label{eq:llvielbeinansatz}
      &\tau_\mu{}^A =
      \bordermatrix{& - & + \cr
      i & \tau_i & \ell\,\alpha_i \cr
      u & 0& \ell }\,,
      && e_\mu{}^{A'} =
      \bordermatrix{& A' \cr
      i & e_i{}^{A'} \cr
      u & 0}\,,
      && b_{ui} = \beta_i\,, && b_{ij} = \beta_{ij} + 2\,\alpha_{[i}\beta_{j]}\,, &&\rme^\phi = \ell\,\rme^\varphi\,,
  \end{align}
  where $\tau_i$, $\alpha_i$, $\ell$, $e_i{}^{A'}$, $\beta_i$, $\beta_{ij}$ and $\varphi$ are all $u$-independent. Taking this as a reduction Ansatz, we find that the remaining unfixed DSNC$^-$ symmetry transformation rules \eqref{eq:symmsDSNCm} reduce, upon using \eqref{eq:llvielbeinansatz}, to the following nine-dimensional ones:
  \begin{align}\label{eq:symmetriesllred}
        &\delta\tau_i= (\lambda_M + \lambda_D) \tau_i \,,&&\delta\phi = \lambda_M\,, \nonumber \\
        &\delta \alpha_i=\partial_i \alpha  \,,&& \delta \beta_i = \partial_i\beta\,, \nonumber \\
        &\delta e_{i}{}^{A'} =\lambda^{A'}{}_{B'}e_{i}{}^{B'}+\lambda^{A'}\tau_{i}\,, && \delta \beta_{ij} = 2\,\partial_{[i}\theta_{j]} - 2\,\alpha_{[i} \partial_{j]} \beta + 2\,\lambda_{A'}\,\tau_{[i} e_{j]}{}^{A'}\,, \nonumber \\
    &\delta \ell = (\lambda_D - \lambda_M) \ell \,,
  \end{align}
  where we have denoted $\lambda_A' = -\lambda_{- A'}$, $\alpha = \xi^u$ and $\beta = - \theta_u$ and where we have not written down the transformations under nine-dimensional diffeomorphisms with parameters $\xi^i$, since they are standard. There are again three vectors $(\tau_i,\alpha_i,\beta_i)$ in the nine-dimensional theory and all three of them are now invariant under the unbroken boosts with parameters $\lambda_{A'}$. The field content is reminiscent of Newton-Cartan structures, but differs in one crucial aspect: there is no analogue of the central charge gauge field $m_i$ transforming under Galilean boosts. Instead, there is a two-form $\beta_{ij}$ which transforms under boosts and should thus be seen as part of the geometry---putting the above somewhere between particle and string Newton-Cartan geometries. As far as we know, this exotic non-Lorentzian geometry has not appeared in the literature so far. It would be interesting to further study the structure of this geometry and understand how it couples to matter.

  If we consider a second DSNC$^-$ theory, whose fields and transformation parameters are denoted with a tilde and reduce via an analogous Ansatz (with tildes), one is led to nine-dimensional transformation rules that are obtained by putting tildes on all fields and parameters of \eqref{eq:symmetriesllred}. The map specified by
  \begin{align}
      &\tilde\tau_i = \tau_i\,, && \tilde \ell = \ell^{-1}\,, &&\tilde{e}_{i}{}^{A'} = e_{i}{}^{A'}\,, \notag\\
      & \tilde \alpha_i = \beta_i\,, && \tilde \beta_i = \alpha_i\,&&\rme^{\tilde\varphi} = \ell\,\rme^{\varphi}\,,\notag\\
      & \tilde \beta_{ij} = \beta_{ij} + 2\,\alpha_{[i}\beta_{j]}\,,
  \end{align}
  and
  \begin{align}
    &\tilde{\alpha} = \beta \,, && \tilde\beta = \alpha\,, && \tilde{\lambda}_M = \lambda_D \,, \notag\\ & \tilde{\lambda}_D = \lambda_M \,, &&\tilde{\theta}_i = \theta_i \,, && \tilde{\lambda}_{A'} = \lambda_{A'} \,,
  \end{align}
  maps the transformation rules with tilde to those without tilde and vice versa. This map constitutes the lightlike T-duality rules (\ref{eq:TDualitiesLL}\hskip -.13truecm) from the nine-dimensional point of view. This can be checked explicitly by substituting the Ansatz \eqref{eq:llvielbeinansatz} (and its tilded version) into the lightlike T-duality rules, that were derived as a worldsheet duality in \cite{Bergshoeff:2018yvt}.

  \section{Supersymmetric Longitudinal T-duality}\label{sec:susyTduality}

\noindent In this section we will address the question whether longitudinal T-duality can be extended to include target space supersymmetry. In other words, we are going to address the question  whether there is a T-duality relation between the minimal non-relativistic supergravity multiplet presented in \cite{Bergshoeff:2021tfn} and $\mathcal N=(1,0)$ supergravity with a null isometry. It is not a priori clear that such a relation is viable since the presence of a lightlike Killing vector $k^\mu \partial_\mu=\partial_z$ is not naturally invariant under supersymmetry. Indeed, in the presence of such a Killing vector, one has that $\tilde{G}_{zz} = 0$ identically, while
  \begin{align}\label{eq:k^2var}
      \delta_\varepsilon\tilde G_{zz} = \bar{\tilde\varepsilon}\,\Gamma_+\tilde\Psi_z \neq 0\,.
  \end{align}
To maintain supersymmetry, we thus have to constrain $\tilde{\Psi}_z$ so that the right-hand-side of \eqref{eq:k^2var} is zero. It then has to be checked whether these extra fermionic constraints are themselves consistent with supersymmetry and if not, additional bosonic constraints have to be imposed. Continuing in this way leads to a chain of constraints that vary to each other under supersymmetry. It is not a priori clear whether this procedure leads to an interesting supergravity multiplet with a null isometry or whether it leads to an overconstrained multiplet. Note that no such problem occurs for a spatial Kaluza-Klein reduction. In that case, the metric component along the compactified direction is not constrained to be zero. Instead, it corresponds to a Kaluza-Klein scalar that forms part of a lower-dimensional supermultiplet.

In the next subsection, we will show that the null reduction of the $\mathcal N=(1,0)$ theory is only consistent if we shorten the multiplet by a set of constraints. The resulting multiplet will be referred to as the $\mathcal N=(1,0)_0$ multiplet. In subsection \ref{sec:GOsugra}, we will review the results of \cite{Bergshoeff:2021tfn}, where a supersymmetric extension of DSNC${}^-$ gravity was derived that is based on a supermultiplet, referred to as the DSNC${}^-$ multiplet in the following. We will also give some details on the Killing spinor equations and general properties of supersymmetric solutions of DSNC$^-$ supergravity. Throughout this section, we follow the conventions of \cite{Bergshoeff:2021tfn}.

\subsection[Consistency of the $\mathcal N=(1,0)_0$ Multiplet]{Consistency of the \boldmath $\mathcal N=(1,0)_0$ Multiplet}\label{sec:(1,0)_0}

\noindent The minimal supergravity multiplet in ten dimensions contains the NS fields $(\tilde{G}_{\mu\nu},\tilde{B}_{\mu\nu},\tilde{\Phi})$, a gravitino $\tilde{\Psi}_\mu$, and a dilatino $\tilde{\Lambda}$. It realizes $16$ supercharges and we take the Majorana-Weyl supersymmetry parameter $\tilde{\varepsilon}$ to be left-handed---i.e., satisfying $\Gamma_{11}\tilde\varepsilon = \tilde{\varepsilon}$---without loss of generality. For this reason we refer to this multiplet as the $\mathcal N=(1,0)$ multiplet. At lowest order in fermions the supersymmetry rules read as follows
  \begin{align}
      &\delta {\tilde E}_\mu{}^{\hat A}  = \,\bar{\tilde{\varepsilon}}\,\Gamma^{\hat A}\tilde{\Psi}_{\mu}\,, && \delta\,\tilde{B}_{\mu\nu} = 2\,\bar{\tilde{\varepsilon}}\,\Gamma_{[\mu}\tilde{\Psi}_{\nu]}\,,&&\delta\,\tilde{\Phi} = \frac12\,\bar{\tilde{\varepsilon}}\,\tilde{\Lambda}\,,\notag\\
      &\delta\,\tilde{\Psi}_\mu = \tilde{D}_\mu^{(+)}\tilde{\varepsilon}\,,&& \delta\,\tilde{\Lambda} = \Big(\Gamma^\mu\,\partial_\mu\tilde{\Phi} - \frac{1}{12}\,\Gamma^{\mu\nu\rho}  \tilde{\mathcal H}_{\mu\nu\rho}\Big)\tilde{\varepsilon}\,,
  \end{align}
where $\tilde{\mathcal H}_{\mu\nu\rho} = 3 \partial_{[\mu} \tilde{B}_{\nu\rho]}$ and $\Gamma_\mu = \tilde E_\mu{}^{\hat A}\Gamma_{\hat A}$. The covariant derivative $\tilde{D}_\mu^{(+)}$ is defined with respect to a torsionful spin connection
  \begin{align}
  \tilde{D}_\mu^{(+)}\tilde{\varepsilon} = \partial_\mu\tilde{\varepsilon} - \frac14 \tilde{\Omega}_\mu{}^{(+)\hat A \hat B} \Gamma_{\hat{A}\hat{B}} \tilde{\varepsilon} \,, \qquad \text{with } \ \ \  \tilde{\Omega}_\mu{}^{(+)\hat A\hat B} = \tilde{\Omega}_\mu{}^{\hat A\hat B} +\frac12\,\tilde{\mathcal H}_\mu{}^{\hat A\hat B}\,,
  \end{align}
where $\tilde{\Omega}_\mu{}^{\hat{A}\hat{B}}$ is the usual Levi-Civita spin connection. It is worth mentioning that this multiplet is not satisfactory for applications in superstring theory. First, it has gravitational anomalies \cite{Alvarez-Gaume:1983ihn}, which render the theory quantum mechanically inconsistent. These can be lifted by coupling to an appropriate Yang-Mills multiplet---leading to either type I or heterotic supergravity. Secondly, it is known that $\mathcal N=(1,0)$ supergravity does not describe the low-energy dynamics of any superstring theory. For these two reasons, we should really see the study of this multiplet as an intermediate step in understanding the full type I, IIA/B, or heterotic supergravity. Phrased differently, we are studying the common sector of all supergravity theories in ten dimensions.

  Let us assume that the theory that we consider has a lightlike isometry with Killing vector $k^\mu \partial_\mu=\partial_z$. Just as in the bosonic sector this implies that the supersymmetry parameter, the gravitino and the dilatino are independent of the isometry $z$-direction. Similar to how the bosonic fields can be parametrized via a lightlike reduction Ansatz \eqref{eq:vielbeinansatz}, it is then useful to decompose the fermions as
  \begin{align} \label{eq:fermlldecomp}
      &\tilde{\varepsilon}=\tilde{\varepsilon}_+ + \tilde{\varepsilon}_-\,, && \tilde{\Psi}_z = \tilde{\psi}_+ + \tilde{\psi}_-\,,&& \tilde{\Psi}_i -\tilde{E}_i{}^+\,\tilde{\Psi}_z = \tilde{\psi}_{i+} + \tilde{\psi}_{i-}\,, && \tilde{\Lambda} =\tilde{\lambda}_+ + \tilde{\lambda}_-\,,
  \end{align}
  where the sublabel $+/-$ refers to the eigenvalue under $\Gamma_{+-}=-\Gamma_{01}$, e.g., $\Gamma_{+-}\tilde\psi_- = -\tilde\psi_-$. Note that then $\Gamma_\pm\tilde\psi_\mp =0$ and that the label $\pm$ of the fermions should not be confused with a flat longitudinal index in lightcone coordinates. The decomposition \eqref{eq:fermlldecomp} is chosen such that the lower-case expressions correspond to non-relativistic nine-dimensional fermions, upon reducing over the lightlike isometry. This is most easily seen by considering the reduced transformation rules under the unfixed boosts $\tilde{\lambda}_i = \tilde{e}_i{}^{A'}\tilde{\Lambda}_{A'-}$:
  \begin{align}
      &\delta\tilde{\psi}_+ = 0\,, && \delta \tilde{\psi}_- = \frac12\,\tilde{\lambda}^{A'+}\Gamma_{A'+}\tilde{\psi}_+\,,
  \end{align}
  and analogously for the other spinor fields. We recognize these as Galilean-type boost transformation rules for spinors. As before, the parametrization is chosen so that all the fields are invariant under the diffeomorphisms in the isometry direction $\tilde\xi^z = -\beta$.

  Let us now turn to the central question of this subsection, namely whether one can consistently impose a null isometry on this multiplet. Using that $\Gamma_z = {\tilde E}_z{}^{\hat A} \Gamma_{\hat A} = \Gamma_+$, we  deduce from eq.~\eqref{eq:k^2var} that we are forced to impose the fermionic constraint $\tilde{\psi}_+=0$. To analyze the supersymmetric consequences of this constraint, it is convenient to write down the supersymmetry variation of both $\tilde{\psi}_+$ and $\tilde{\psi}_-$:
  \begin{align}
      \delta\tilde{\psi}_+ &= -\frac14\,\tilde{Z}^{A'B'}\Gamma_{A'B'}\tilde{\varepsilon}_+\,,\label{first}\\
      \delta\tilde{\psi}_- &= -\frac14\,\tilde{Z}^{A'B'}\Gamma_{A'B'}\tilde{\varepsilon}_- + \frac12\,\tilde{Z}^{A'0}\Gamma_{A'+}\tilde{\varepsilon}_+\,.\label{second}
  \end{align}
  We have used here some results on lightlike reductions  \cite{Bergshoeff:2017dqq} and the fact that, using eq.~\eqref{eq:vielbeinansatz},
  $\tilde{\mathcal H}_{zij} = -2\,\partial_{[i}\tilde{n}_{j]}$. We have also denoted
  \begin{equation}
  \tilde{Z}_{A'B'} = \tilde{e}_{A'}{}^i \tilde{e}_{B'}{}^j \tilde{Z}_{ij}\,,\hskip 2.5truecm \tilde{Z}^{A'0}=\tilde{e}^{A'i}\tilde{\tau}^j\,\tilde{Z}_{ij}\,,
  \end{equation}
   where $\tilde{Z}_{ij}$ is defined in \eqref{eq:NS0constraint}. Clearly, the transformation rule  \eqref{first} shows that the constraint $\tilde{\psi}_+=0$ is not invariant and leads to the constraint $\tilde{Z}_{A'B'}=0$.
  Using that
  \begin{equation}
  \delta \tilde{Z}_i = 2\,\bar{\tilde{\varepsilon}}_+\,\Gamma_i\tilde{\psi}_- + 2\,\bar{\tilde{\varepsilon}}_-\Gamma_i\tilde{\psi}_+\,,
  \end{equation}
   we can schematically give the supersymmetry consequences of $\tilde{G}_{zz}=0$ up to the third iteration (and ignoring higher order terms in fermions) as follows
  \begin{align} \label{eq:varsequence}
  \tilde{G}_{zz}=0 \quad\overset{\delta_{\tilde{\varepsilon}}}{\longrightarrow} \quad\tilde{\psi}_+ = 0 \quad \overset{\delta_{\tilde{\varepsilon}}}{\longrightarrow}\quad \tilde{Z}_{A'B'}=0\quad \overset{\delta_{\tilde{\varepsilon}}}{\longrightarrow} \quad \tilde{e}_{A'}{}^i \tilde{e}_{B'}{}^j\,\partial_{[i}(\bar{\tilde{\varepsilon}}_+\Gamma_{j]}\tilde{\psi}_-)=0\,.
  \end{align}
  At this point, we see that the last constraint is a differential constraint on $\tilde{\psi}_-$ and $\tilde{\varepsilon}_+$. Instead of continuing to vary these constraints, we note that the sequence \eqref{eq:varsequence} terminates if $\tilde{\psi}_-=0$. This in turn implies that $\delta \tilde{Z}_i =0$ and also, according to the transformation rule \eqref{second}, that  $\tilde{Z}^{A'0}=0$. This leads to the following supersymmetric set of constraints on the supergravity fields:
  \begin{align}\label{eq:(1,0)constraints}
      &\tilde{G}_{zz}=0\,, && \tilde{\psi}_+ = 0\,,&& \tilde{\psi}_- =0\,, && \tilde{Z}_{ij}=0\,,
  \end{align}
  or, phrased more covariantly:
  \begin{equation}
  \{k^\mu k_\mu\,,k^\mu\tilde{\Psi}_\mu\,,\partial_{[\mu}\tilde{Z}_{\nu]}\}=0\hskip  1truecm \textrm{with}\hskip 1truecm \tilde{Z}_\mu = k^\nu(\tilde{G}_{\nu\mu}-\tilde{B}_{\nu\mu})\,.
  \end{equation}
  The multiplet of constraints \eqref{eq:(1,0)constraints} effectively shortens the $\mathcal N=(1,0)$ multiplet by $\mathbf 1$ bosonic algebraic, $\mathbf{16}$ fermionic algebraic, and $\mathbf{36}$ bosonic differential constraints. We refer to this constrained multiplet as the $\mathcal N=(1,0)_0$ multiplet. Of course, this construction only makes sense if all fields and parameters are independent of the $z$-coordinate, which means that the theory is effectively nine-dimensional.

\subsection{Review of the DSNC${}^-$ Multiplet}\label{sec:GOsugra}

\noindent In this section, we review the essential results of \cite{Bergshoeff:2021tfn}, where the first example of non-relativistic supergravity theory in ten dimensions was presented, namely a supersymmetric extension of DSNC${}^-$ gravity that arises from a stringy non-relativistic limit of $\mathcal N=(1,0)$ supergravity. This limit is quite subtle in that it requires dealing with terms that diverge in the limit. Here we will only give the final results and central properties and refer the reader to \cite{Bergshoeff:2021tfn} for more details.

The fields of the DSNC$^-$ supergravity multiplet   consist of the bosonic fields $\tau_\mu{}^{A}$, $e_\mu{}^{A'}$, $b_{\mu\nu}$ and $\phi$ of DSNC$^-$ gravity, as well as gravitini $\psi_{\mu \pm}$ and dilatini $\lambda_\pm$ \cite{Bergshoeff:2021tfn}. This multiplet has the same number of field components as the relativistic $\mathcal N=(1,0)$ multiplet, and is related to it via a contraction parameter $\omega$ as follows
\begin{align}
  &\tau_\mu{}^A = \omega^{-1}\,\tilde E_\mu{}^A\,, && e_\mu{}^{A'}=\tilde E_\mu{}^{A'}\,, && \rme^\phi = \omega^{-1}\,\rme^{\tilde\Phi}\,,\notag\\
  &b_{\mu\nu} = \tilde B_{\mu\nu} + \tilde E_\mu{}^A \tilde E_\nu{}^B\,\epsilon_{AB}\,, && \psi_{\mu\pm}=\omega^{\mp 1/2} \Pi_{\pm}\tilde\Psi_\mu\,, && \lambda_\pm = \omega^{\mp 1/2}\Pi_{\pm}\tilde\Lambda\,.
\end{align}
This particular re-scaling guarantees that the fermions $\psi_{\mu\pm}$ and $\lambda_\pm$ form a faithful representation of Galilean boosts as $\omega\to\infty$. The other bosonic symmetries follow analogously. However, it turns out that there is a non-zero term at order $\mathcal O(\omega^2)$ in the expansion of the supersymmetry rules. Naively this implies that the non-relativistic limit $\omega\to\infty$ is ill-defined. In \cite{Bergshoeff:2021tfn} we have shown that this inconsistency can be lifted by
\begin{enumerate}
  \item Imposing a constraint on the intrinsic torsion of the geometry $T_{\mu\nu}^\rho\tau_\rho{}^-=0$, or
  \begin{align}
    \tau^-\wedge\rmd\tau^-=0\,,
  \end{align}
  which corresponds to $\mathbf{36}$ independent first order differential constraints on the longitudinal Vielbein $\tau_\mu{}^-$. Alternatively, it can be expressed as $\tau_{A'B'}{}^-=0$ and $\tau_{A'}{}^{--}=0$. This constraint is consistent with all the symmetries, in particular with supersymmetry since $\delta_\epsilon \tau_\mu{}^- = 0$.
  \item Including additional symmetries $\delta_D(\lambda_D),\,\delta_S(\eta_-)$, and $\delta_T(\rho_-)$, which we refer to as anisotropic dilatations and fermionic $S$- and $T$-symmetries, respectively. All of these additional symmetries act as local shifts on some fields
  \begin{align}
    &\delta \tau_\mu{}^A = \lambda_D\tau_\mu{}^A\,,&& \delta \phi = \lambda_D\,,\notag\\
    &\delta \psi_{\mu+} = \frac12 \,\lambda_D\psi_{\mu+} + \frac12\,\tau_\mu{}^+\Gamma_+\eta_-\,, && \delta\psi_{\mu-} = -\frac12\,\lambda_D\psi_{\mu-} + \frac12\,\tau_\mu{}^+\rho_-\,, \notag\\
    &\delta \lambda_+ = \frac12\,\lambda_D\lambda_+\,, && \delta \lambda_- = -\frac12\,\lambda_D\lambda_- + \eta_-\,. \label{eq:ST2}
  \end{align}
  These extra symmetries give $\mathbf{1}$ bosonic and $\mathbf{8}+\mathbf{8}$ fermionic gauge symmetries. They are absent in the relativistic parent theory and imply that the multiplet is effectively smaller since the local shift symmetries can be gauge fixed by setting some field components to constant values.
\end{enumerate}
The fact that the divergent limit can be regularized and made sense of is highly non-trivial and a genuine feature of ten-dimensional supergravity. We have shown in \cite{Bergshoeff:2021tfn} that the same limit is well-defined at the level of the equations of motion---which are conjectured to capture the universal sector of the beta functions of non-relativistic superstring theory.

It turns out  that the emergent symmetries of the DSNC${}^-$ multiplet  are in one-to-one correspondence to the multiplet of constraints \eqref{eq:(1,0)constraints} in the $\mathcal N=(1,0)_0$ theory. More precisely, the emergent  gauge symmetries of the DSNC${}^-$ multiplet are T-dual to the constraints we found for the NS${}_0$ multiplet:
\begin{align}
    & && \mathrm{DSNC}^-~\mathrm{supergravity} && \mathcal N=(1,0)_0~\mathrm{supergravity}\notag\\
    &\mathbf 1: && \delta_D && \tilde G_{zz}=0\,,\notag\\
    &\mathbf{8+8}: && \delta_S + \delta_T && \tilde\Psi_z = 0\,,\notag\\
    &\mathbf{36}: && \tau^-\wedge\rmd \tau^-=0\,,&& \tilde Z_{ij}=0\,.
\end{align}
These observations indicate that the T-duality between the DSNC${}^-$ and the $\mathcal N=(1,0)_0$ multiplets can be extended beyond the bosonic sector. We will give some details about the T-duality rules for the fermions below.

In the next section, we will discuss half-supersymmetric solutions of DSNC$^-$ supergravity. For this, we need the Killing spinor equations of this supergravity theory that follow from setting all the fermions of the theory to zero and requiring by consistency that $\delta_{\epsilon}\lambda_\pm=0$ and $\delta_{\epsilon}\psi_{\mu\pm}=0$. These equations have been derived in \cite{Bergshoeff:2021tfn} and read
\begin{align}\label{eq:KSEs}
0 &=\Gamma^{A'}\epsilon_{-}\mathcal D_{A'}\phi-\frac{1}{12}\,h_{A'B'C'}\Gamma^{A'B'C'}\epsilon_{-}+\eta_-\,,\notag\\
0 &=\Gamma^{A'}\epsilon_{+}\mathcal D_{A'}\phi-\frac{1}{12}\,h_{A'B'C'}\Gamma^{A'B'C'}\epsilon_{+}+\frac{1}{2}\,T^{A'B'}\Gamma_{A'B'+}\epsilon_{-}\,,\notag\\
\mathcal D_{\mu}\epsilon_{-}-\frac12\,\omega_{\mu}{}^{-A'}\Gamma_{-A'}\epsilon_+ &= \frac{1}{8}\,e_{\mu}{}^{C'}h_{A'B'C'}\Gamma^{A'B'}\epsilon_{-} - \tau_{\mu}{}^+ \rho_-\,,\notag\\
\mathcal D_{\mu}\epsilon_{+}&=\frac{1}{8}\,e_{\mu}{}^{C'}h_{A'B'C'}\Gamma^{A'B'}\epsilon_{+}-T_\mu{}^{A'}\Gamma_{A'+}\epsilon_{-}-\frac12\,\tau_{\mu}{}^+\Gamma_{+}\eta_-\,,
\end{align}
where $\mathcal D_\mu$ is a $\mathsf{SO}(1,1)\times\mathsf{SO}(8)$ and dilatation covariant derivative and we have furthermore defined $T_\mu{}^{A'} = e_{\mu}{}^{B'}\tau{}_{B'}{}^{A'+}+\tau_{\mu}{}^-\tau^{A'++}$. We refer to appendix \ref{sec:TSNC} and \cite{Bergshoeff:2021tfn} for more details on our notation and conventions. Note that these Killing spinor equations are consistent with the bosonic symmetries of the theory. In particular, one can check that the equations form a non-trivial reducible but indecomposable representation under Galilean boosts.\footnote{This supposes that one also assigns a boost transformation rule to the Killing spinors $\epsilon_\pm$ that is analogous to the boost transformation rules of the fermionic fields of DSNC$^-$ supergravity.}

In this paper, we will be mostly interested in half-supersymmetric solutions---i.e., solutions that preserve $8$ of the $16$ supersymmetries $(\epsilon_+,\epsilon_-)$. The $S-$/$T-$parameters $(\eta_-,\rho_-)$ are Killing spinors too. However, we do not include them when counting the preserved supercharges, since they are just an artifact of the initial overparametrization as explained above. Relatedly, it is clear from \eqref{eq:KSEs} that solutions $(\eta_-,\rho_-)$ exist independent of the background geometry. In this paper, we will not attempt to solve the Killing spinor equations systematically and instead refer the reader to \cite{Bergshoeff:2020baa} where analogous Killing spinor equations for a Galilean system in three dimensions have been studied.

\section{The Half-supersymmetric NS Branes}\label{sec:susyBranes}

\begin{figure}[t]
\centering
\includegraphics[width=.7\textwidth]{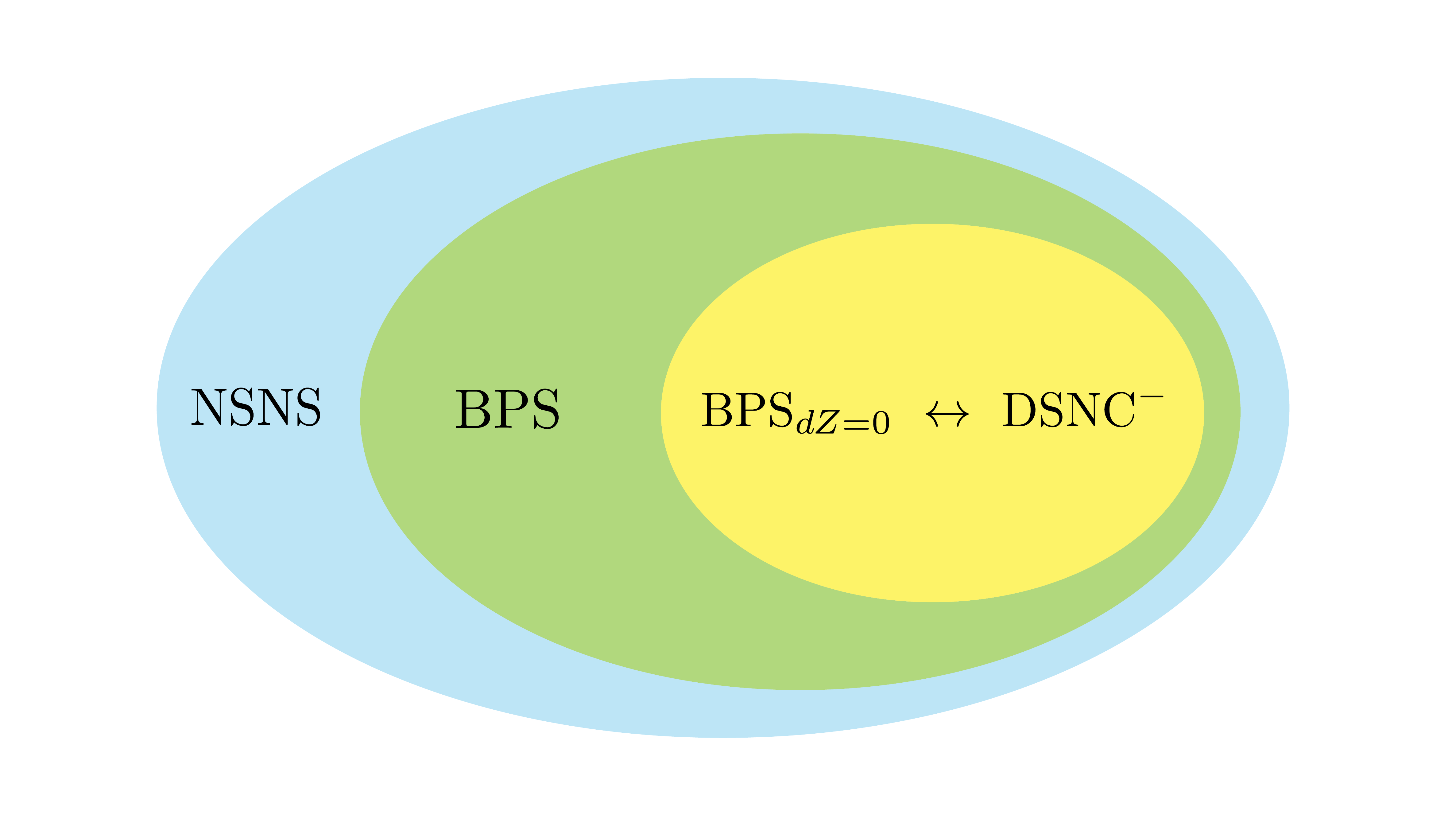}
\caption{A schematic depiction of the types of solutions considered in this section. The blue outline represents all the solutions of pure NS gravity. The green outline represents the solutions that preserve at least one supercharge. This implies that the solution has a lightlike Killing vector \cite{Fontanella:2019avn}. The yellow outline denotes the solutions of $\mathcal N=(1,0)_0$ supergravity, i.e.~those that can be mapped to solutions of the DSNC${}^-$ supergravity theory. The boundary between the green and the yellow solutions is delineated by the constraint \eqref{eq:NS0constraint}.}
\label{fig:blobs}
\end{figure}

\noindent To construct the basic NS brane solutions, we will use the longitudinal T-duality rules \eqref{eq:BuscherRescal}, applied to solutions of $\mathcal{N}=(1,0)_0$ supergravity, to generate solutions of DSNC$^-$ supergravity. One has to be careful in applying longitudinal T-duality blindly to brane solutions of relativistic $\mathcal{N}=(1,0)$ supergravity, for two reasons. Firstly, the T-duality only makes sense if the solution has a lightlike isometry. If one restricts oneself to supersymmetric solutions, this requirement is automatically fulfilled since every solution that preserves at least one supercharge also has a lightlike isometry \cite{Fontanella:2019avn}. This implies that all the supersymmetric solutions of $\mathcal N=(1,0)$ supergravity qualify as candidates for the solution generating T-duality. Secondly, the set of solutions of $\mathcal N=(1,0)_0$ supergravity is not equivalent to the set of solutions of $\mathcal N=(1,0)$ supergravity that have a lightlike isometry since solutions of $\mathcal N=(1,0)_0$ supergravity also need to satisfy the bosonic constraint \eqref{eq:NS0constraint}. In particular, there exist supersymmetric solutions of the $\mathcal N=(1,0)$ theory that are not solutions of the $\mathcal{N}=(1,0)_0$ theory. We have summarized the situation in figure \ref{fig:blobs}.

In subsection \ref{sec:windingstring} we consider the half-supersymmetric pp-wave solution of $\mathcal N=(1,0)$ supergravity. It has a null isometry and furthermore solves the constraint \eqref{eq:NS0constraint}. Hence we can apply longitudinal T-duality to obtain a solution of DSNC${}^-$ supergravity that we refer to as the `winding string' solution. This name will be motivated further by a study of source terms in section \ref{sec:sources}. The solution has vanishing intrinsic torsion and reduces to a conventional NC geometry in nine dimensions.

In subsection \ref{sec:unwoundstring} we consider the half-supersymmetric fundamental and anti-fundamental string solutions of $\mathcal N=(1,0)$ supergravity. Both solutions have a null isometry. However, only the fundamental string solution also solves the constraint \eqref{eq:NS0constraint}. In other words, the fundamental string is a solution of $\mathcal N=(1,0)_0$ supergravity whereas the anti-fundamental string is not. The corresponding background of DSNC${}^-$---obtained via longitudinal T-duality from the fundamental string solution---is referred to as the `unwound string solution'. This name will again be clarified by studying
the structure of the source term in section \ref{sec:sources}. The unwound string solution has non-vanishing intrinsic torsion and reduces to a matter-coupled torsional Newton-Cartan geometry in nine dimensions.
%

It is worth mentioning that the solution generating technique presented is more general. It can be applied to any solution of $\mathcal N=(1,0)_0$ supergravity. To illustrate this, we use the same longitudinal T-duality to also find non-relativistic versions of the NS 5-brane and the Kaluza-Klein (KK5) monopole solutions, thus obtaining all the basic half-supersymmetric non-relativistic NS branes. The consistency of the duality guarantees that it leads to solutions of the DSNC${}^-$ theory. In principle, one can access an infinity of non-trivial solutions---among which also intersecting branes and other exotic objects. We leave that for future work.

\subsection{The Winding String Solution}\label{sec:windingstring}

\noindent To construct the winding string solution of DSNC${}^-$ gravity, we take the well-known pp-wave solution \cite{Gueven:1987ad,Bergshoeff:1992cw} of NS gravity as our starting point:
\begin{align} \label{eq:pp-wave}
&d\tilde{s}_{pp}^2 = - 2\,dt (dz + K\,dt) + dz_{(8)}^{2}\,,&&  \tilde{B}=0\,,&&\rme^{\tilde{\Phi}}= g_s \,.
\end{align}
Here, we have used the ten-dimensional coordinates $x^\mu =(z,x^i)=(z,t,z_{(8)})$, see also appendix \ref{sec:conventions} for our index conventions. Furthermore, $g_s$ is a constant that corresponds to the string coupling constant of the underlying string theory. The profile function $K=K(t,z_{(8)})$ is a harmonic function of the eight transverse coordinates $z_{(8)}$ and can have an arbitrary dependence on $t$.\footnote{For an early study of a pp-wave background T-dual to a non-relativistic AdS${}_5\times$S${}^5$ background, see \cite{Gomis:2005pg}.} This solution is known to be half-supersymmetric, i.e., it preserves $8$ of the 16 supercharges, namely those with constant supersymmetry parameter $\tilde\varepsilon$, satisfying $\Gamma_{01} \tilde\varepsilon=\tilde\varepsilon$.

The pp-wave  solution has a lightlike isometry in the $z$-direction. It also satisfies the constraints \eqref{eq:NS0constraint} and is therefore a solution of NS${}_0$ gravity.\,
In order to map it to a solution of DSNC${}^-$ gravity, it is easiest to apply the inverse of the T-duality rules given in  \eqref{eq:BuscherRescal}. In particular, this T-duality  maps the profile function $K$ of the pp-wave to the worldsheet component $b_{zt}$ of the Kalb-Ramond 2-form. We thus end up with the following solution of DSNC${}^-$ gravity:
\begin{align}\label{eq:NF1}
    &ds^2_\tau= -dt^2 + dz^2\,, && ds^2_e = dz_{(8)}^2\,, && b = K\,dt\wedge dz\,, && \rme^\phi = g_s\,,
\end{align}
where $ds^2_\tau = \eta_{AB}\tau_\mu{}^A\tau_\nu{}^B dx^\mu dx^\nu$ and $ds^2_e = \delta_{A'B'}e_\mu{}^{A'} e_\nu{}^{B'} dx^\mu dx^\nu$. We will refer to eq.~\eqref{eq:NF1} as the winding string solution. Note that this solution is invariant under the NR transverse T-duality rules \eqref{transverseT}.

The above solution \eqref{eq:NF1} indeed describes the gravitational field caused by a string where the profile function $K$ plays the role of the Newton potential. The possible time dependence of $K$ can then be understood as indicating that the string moves in a frame with a time-dependent acceleration. From \eqref{eq:NF1}, one can see that the winding string solution is  sourced by the $w=1$, $n=0$ state of the non-relativistic string spectrum. Indeed, the solution \eqref{eq:NF1} has a non-trivial $b$-field in the world-sheet directions, as one expects from a solution that is sourced by a string that winds around the spatial longitudinal direction. We will confirm this in the next section.

Using the reduction rules given in section \ref{sec:longitudinalTD}, it is straightforward to show that the winding string solution reduces to the following nine-dimensional solution
\begin{align}\label{combination2}
  &\tau_i dx^i = dt\,, && \delta_{A'B'}e_i{}^{A'}e_j{}^{B'}dx^idx^j = dz_{(8)}^2\,,&& m_idx^i = K\,dt\,,&& \rme^\phi = g_s\,.
\end{align}
This torsionless Newton-Cartan geometry describes the usual Newtonian gravitational field outside a massive non-relativistic point particle. We will show how this follows from a double dimensional reduction in section \ref{sec:sources}. The same solution can be obtained by a null reduction of the pp-wave. A similar reduction has been considered in \cite{Bergshoeff:2017vjg}.

Let us now evaluate the Killing spinor equations \eqref{eq:KSEs} on the winding string background and find the Killing spinors $(\epsilon_+,\epsilon_-)$ and $(\eta_-,\rho_-)$. This will determine the amount of supersymmetry that is preserved by this solution. Since we are assuming an isometry in the $z$-direction, we take the Killing spinors to be $z$-independent. Using the geometric data as given in \eqref{eq:NF1}, the Killing spinor equations simplify to
\begin{align}
  &\eta_- = 0\,, &&\partial_\mu\epsilon_+ + \frac12\tau_\mu{}^+\Gamma_+\eta_-=0\,, &&\rho_- = \frac14\,\partial_{A'}K\,\Gamma_-\Gamma^{A'}\epsilon_+\,, \notag\\
  & \partial_m\epsilon_- = 0\,,&&\partial_z\epsilon_- =0\,, && \partial_t\epsilon_- = \frac{\sqrt2}{4}\partial_{A'}K\,\Gamma^{A'}\Gamma_-\epsilon_+\,,
\end{align}
where $\partial_m$ denotes a derivative with respect to the transversal coordinates $z_{(8)}^m$. These equations have a unique solution
\begin{align}
  &\big(\eta_- = 0,\rho_- = 0 \big)\,,&&\big(\epsilon_+=0,\epsilon_- = \mathrm{constant}\big)\,,
\end{align}
which shows that this solution preserves half of the supersymmetries, i.e., $8$ supercharges.

We note that the background \eqref{eq:pp-wave} is just one example of a more general class of pp-wave solutions \cite{Brinkmann1} that are defined by the existence of a covariantly constant null vector. It was established in \cite{Gueven:1987ad} that pp-wave solutions of the lowest order relativistic supergravity extend to solutions of supergravity including all $\alpha'$ corrections. For an extension of the above results and references, see \cite{Bergshoeff:1992cw}. We have seen above that the existence of a covariantly constant null vector leads to a string Newton-Cartan geometry with vanishing intrinsic torsion under longitudinal T-duality. It would be interesting to understand the absence of $\alpha'-$corrections directly from the point of view of non-relativistic string theory.

\subsection{The Unwound String Solution}\label{sec:unwoundstring}

\noindent Our starting point is the following NS fundamental string solution, lying in the $z$-direction:
\begin{align}\label{RF1}
&d\tilde{s}_{F1}^2 = - 2\,H^{-1} dtdz + dz_{(8)}^{2}\,,&& \tilde{B} = (H^{-1} - 1)\,dt\wedge dz\,,&&\rme^{\tilde{\Phi}}= g_s\,H^{-1/2}\,,
\end{align}
where $H=H(z_{(8)})$ is harmonic in the transverse directions $z_{(8)}$ and the constant $g_s$ corresponds to the string coupling constant. This solution is also a solution of NS$_0$ gravity, since it solves the constraints \eqref{eq:NS0constraint}. It preserves $8$ supercharges $\varepsilon(z_{(8)}) = H^{-1/4}\zeta$, where $\zeta$ is constant and satisfies $\Gamma_{01}\zeta = -\zeta$. The anti-fundamental string has the opposite sign for the Kalb-Ramond field and $\Gamma_{01}\zeta = +\zeta$.

T-dualizing along the $z$-isometry direction we obtain the following  solution of DSNC${}^-$ gravity:
\begin{align}\label{eq:unwoundstring}
    &ds_\tau^2 = -H^{-2}\,dt^2 + \big(dz + (H^{-1} - 1)dt\big)^2\,, && ds_e^2 = dz_{(8)}^2\,, && \rme^\phi = g_s\,H^{-1/2}\,.
\end{align}
For the purpose of working out the Killing spinor equations, it is useful to give the following corresponding  set of non-zero longitudinal and transverse Vielbeine:
\begin{align}
\tau_{t}{}^{0}&=H^{-1}\,,&\tau_{z}{}^{1}&=1\,,&\tau_{t}{}^{1}&=H^{-1}-1\,,&e_{\mu}{}^{A'}&=\delta_{\mu}^{A'}\,.
\end{align}
Note that this solution has intrinsic torsion $\tau_{A'}{}^{++}= 1/2\,H^{-1}\partial_{A'}H$. For reasons that will be explained in the next section, we will refer to this solution as the {\sl unwound string solution}. Note that  the solution \eqref{eq:unwoundstring} satisfies the consistency requirement \eqref{constraint2} due to a cancellation of the harmonic functions in the expression for $\tau_\mu{}^-$. This is  a direct consequence of the fact that the fundamental string solution solves the NS${}_0$ constraint \eqref{eq:NS0constraint}.\,\footnote{Conversely, it implies that the unwound anti-string---which one obtains by T-dualizing the anti-fundamental string---does not solve the DSNC${}^-$ constraint \eqref{constraint2}. We see that half of the fundamental string solutions get eliminated by the consistency requirement \eqref{constraint2} or \eqref{eq:NS0constraint}, respectively.}

From a nine-dimensional point of view, the unwound string solution \eqref{eq:unwoundstring} is given by the following solutions for the nine-dimensional matter and torsional Newton-Cartan background fields:
\begin{align} \label{eq:torsional9d}
&\tau_i dx^{i} = - H^{-1} dt\,, && \delta_{A'B'}e_i{}^{A'}e_j{}^{B'}dx^idx^j = dz_{(8)}^2\,,\notag\\
&n_i dx^{i} = (H^{-1}-1)dt\,, && e^\phi = g_s H^{-1/2}\,.
\end{align}
Note that this solution satisfies the geometric constraint \eqref{eq:(1,0)constraints}, i.e., $\partial_{[i}\tau_{j]}=-\partial_{[i}n_{j]}$ and that the resulting Newton-Cartan geometry is twistless torsional, i.e., $\tau_{[i} \partial_j \tau_{k]} = 0$.

Finally, let us determine the Killing spinors of the unwound string solution. Using the fact that the dilaton is covariantly constant, it is straightforward to evaluate the Killing spinor  eqs.~\eqref{eq:KSEs} on the geometric data \eqref{eq:unwoundstring} to find
\begin{align}
    & \eta_- = 0 \,, \qquad \qquad \qquad \partial_\mu\epsilon_- -\frac12 \left(H^{-1} \partial_\mu H \right) \, \epsilon_- + \tau_\mu{}^+\rho_-=0\,, \nonumber \\ & \partial_\mu\epsilon_+ + \frac12 \left(H^{-1} \partial_\mu H \right) \, \epsilon_+  + \frac12 \tau_\mu{}^- H^{-1} \partial_{A'} H \Gamma^{A'}{}_{+}\epsilon_-=0\,.
\end{align}
These equations have  a unique solution corresponding to $8$ preserved supercharges
\begin{align}
   &\big(\eta_-=0,\rho_-=0)\,, && \big(H^{1/2} \epsilon_+=\mathrm{constant},\epsilon_-=0)\,.
\end{align}

We have derived two half-supersymmetric solutions of the DSNC${}^-$ theory: the winding string \eqref{eq:NF1} and the unwound string \eqref{eq:unwoundstring}. Let us now ask whether these two backgrounds are related to each other by an appropriate notion of T-duality. We will see that this is, indeed, the case (as long as the profile function in the winding string background is time-independent $K=K(z_{(8)})$). In fact, it is natural to expect such a relation since the associated relativistic solutions---the relativistic fundamental string solutions \eqref{RF1} and the pp wave \eqref{eq:pp-wave}---are T-dual along a spatial isometry $\partial_y = 2^{-1/2}(\partial_z-\partial_t)$. They are mapped to each other via the Buscher rules \eqref{eq:Buscher}. Hence one can, at least in principle, establish a map between the two solutions by composing a longitudinal T-duality with a spatial and an inverse longitudinal one: $T(\partial_z)\circ T(\partial_y)\circ T^{-1}(\partial_z)$, where the expression in the bracket indicates the respective isometry. Instead of doing this we note that the unwound string solution \eqref{eq:unwoundstring} has a lightlike isometry $\ell = \partial_u = 2^{-1/2}(\partial_z+\partial_t)$ in the sense of eqs. \eqref{eq:DSNClightlike}. Hence, we can apply the lightlike T-duality as explained in section \ref{sec:lightlikeTD}. It is convenient to fix part of the dilatation and $\mathsf{SO}(1,1)$ invariance by choosing $\tau_u{}^+ = H^{-1/2}$ and $\rme^\varphi = g_s$. The other nine-dimensional fields in eq. \eqref{eq:llvielbeinansatz} follow accordingly
\begin{align}
    &\tau_i dx^i = -H^{-1/2}\,dv\,, &&\alpha_i dx^i = (H-1)\,dv &&e_{ij}dx^idx^j = dz_{(8)}^2\,, && \rme^\varphi = g_s\,,
\end{align}
where $v = 2^{-1/2}(z-t)$. Using the T-duality rules  \eqref{eq:llvielbeinansatz} one can then see that we exactly reproduce the winding string solution \eqref{eq:NF1}, if $K=H-1$. This establishes the observation that the time-independent winding string and the unwound string solution are related by lightlike T-duality. These examples also illustrate a more general property of the lightlike T-duality: it can relate backgrounds with vanishing intrinsic torsion to backgrounds with non-vanishing intrinsic torsion.

\subsection{NS 5-Branes and KK5 Monopoles}

\noindent For completeness, we now consider the solitonic NS five-brane solutions of the relativistic theory, namely the NS 5-brane and the KK5 monopole. They are related by electromagnetic duality to the fundamental string  and pp-wave solution, respectively, and to each other by T-duality along a spatial isometry. Moreover, both solutions preserve $8$ supercharges, have a lightlike isometry, and solve the geometric constraints \eqref{eq:(1,0)constraints}. In other words, the solitonic $5$-branes are solutions of the $\mathcal N=(1,0)_0$ theory. As a consequence, we can use longitudinal T-duality to map the NS 5-brane and the KK5 monopole to analogous non-relativistic NS 5-brane and KK5 solutions of the DSNC${}^-$ theory, respectively.

\paragraph{The non-relativistic NS 5-brane:}
The relativistic NS 5-brane solution is given by \cite{Callan:1991dj, Callan:1991at}
\begin{align}\label{eq:NS5}
    &d\tilde{s}_{\rm NS5}^2 = 2\,dtdz + dy_{(4)}^2 + F\, dz^2_{(4)}\,,&&\tilde{\mathcal{H}}_{mnp} =  \varepsilon_{mnp}{}^qF\partial_q F\,, &&\mathrm{e}^{\tilde{\Phi}} = g_s\,F^{1/2}\,,
\end{align}
where $g_s$ is the string coupling constant and we have split the ten-dimensional coordinates $x^\mu$ as $x^\mu = (t, z, y_{(4)}, z_{(4)})$. The function $F=F(z_{(4)})$ is harmonic with respect to the 4 transverse coordinates $z_{(4)}$. Furthermore, it is understood that $\varepsilon_{mnpq}$ refers to the four transverse directions $z_{(4)}$ with $\varepsilon_{\underline{6789}}=1$ where numerical underlined values refer to values taken by curved indices. One can T-dualize the NS 5-brane solution either along a longitudinal spatial isometry direction $\partial_t + \partial_z$ (where $\partial_t$/$\partial_z$ are lightlike) or along a transverse isometry direction $\partial/\partial z_{(4)}^9$ (without loss of generality).\,\footnote{To create a transverse isometry direction one should smear the NS 5-brane, i.e.~consider a periodic array of NS 5-branes in the $z^9$-direction, so that $F = F(z^6, z^7, z^8)$.} T-duality along a longitudinal direction  maps the NS 5-brane onto itself, whereas T-duality along a transverse direction maps the NS 5-brane onto the Kaluza-Klein monopole that we will discuss below.

In this work, however, we are interested in dualizing along the lightlike isometry $\partial_z$. Applying the longitudinal T-duality rules \eqref{eq:BuscherRescal} leads to a valid solution of the DSNC${}^-$ theory, the non-relativistic NS 5-brane:
\begin{align}\label{eq:NRNS5}
    &ds_\tau^2 = -dt^2 + dz^2\,&& ds_e^2 =  dy_{(4)}^2 + F\, dz^2_{(4)} && h_{mnp} = \varepsilon_{mnp}{}^{q} F\partial_{q}F\,, && \mathrm{e}^{\phi} = g_s\,F^{1/2}\,.
\end{align}
This solution is torsionless. It can be checked that it solves the Killing spinor equations \eqref{eq:KSEs}, with zero $\eta_-$ and $\rho_-$ and constant $\epsilon_\pm$, obeying
\begin{align}
\Big(1+\Gamma^{6789}\Big)\epsilon_{\pm}&=0\ .
\end{align}
The non-relativistic NS 5-brane \eqref{eq:NRNS5} is thus half-supersymmetric.

\paragraph{The non-relativistic KK5 monopole:} The relativistic KK5 monopole solution is given by \cite{Gross:1983hb, Sorkin:1983ns}
\begin{align}\label{eq:KK5}
    d\tilde{s}^2_{\rm KK5} = & 2\,dtdz + dy_{(4)}^2 + F\,dz_{(3)}^2 + F^{-1}\left(dy + A_m\, dz^m_{(3)}\right)^2\,,
\end{align}
where $F = F(z_{(3)})$ is harmonic with respect to the three transverse coordinates $z_{(3)}$, and $A_m=A_m(z_{(3)})$ satisfies $2\,\partial_{[m}A_{n]} = \varepsilon_{mnp}\partial^p F$, where $\varepsilon_{\underline{789}}=1$. We have split the ten-dimensional coordinates $x^\mu$ as $x^\mu = (t,z, y_{(4)},y, z^m_{(3)})$ where $(\partial_t + \partial_z,\partial/\partial y_{(4)})$ are the 5 spatial worldvolume isometry directions and $\partial_y$ is a so-called special isometry direction used in the definition of the KK5 solution. This special isometry direction distinguishes the KK5 monopole from a brane solution. The metric is nonsingular at $z^i_{(3)}=0$ if and only if the coordinate $y$ parametrizes a circle with  period $y \sim y + 2\pi$. In order to map this solution to a solution of the DSNC${}^-$ theory we dualize along the $\partial_z$ direction, yielding: the non-relativistic KK5 brane
\begin{align}\label{eq:NRKK5}
   ds_\tau^2 = -dt^2 + dz^2\,, && ds_e^2 = F\,dz_{(3)}^2 + F^{-1}\big(dy + A_m\,dz^m_{(3)}\big)^2\,,&& b = 0\,, && \rme^\phi = g_s\,.
\end{align}
It is immediate to verify that this is a zero torsion solution and it can be checked that it solves the Killing spinor equations \eqref{eq:KSEs} with zero $\eta_-$ and $\rho_-$ and constant $\epsilon_\pm$, obeying
\begin{align}
\Big(1-\Gamma^{6789}\Big)\epsilon_{\pm}&=0\,.
\end{align}
The non-relativistic KK5 monopole is thus also half-supersymmetric.

We note that a T-duality of the relativistic KK5 monopole along the special transverse $\partial_y$ isometry direction maps it to the smeared relativistic NS 5-brane. Similarly, the NR NS 5-brane and KK5 solution are T-dual to each other when applying the transverse Buscher rules \eqref{transverseT} along a transverse direction of the (smeared) NR NS 5-brane or the special $\partial_y$ isometry direction of the NR KK5 monopole. Finally, one could also consider a further T-duality of the relativistic KK5 monopole in one of the three transverse directions. This leads to a family of so-called exotic branes with decreasing number of transverse directions. It is suggestive that similar exotic branes exist in NR string theory, but we will not discuss this further here.

\section{The NS Source Terms}\label{sec:sources}

\noindent In this section, we consider the spectrum of both non-relativistic string theory and the discrete lightcone quantization (DLCQ) of relativistic string theory. The goal is to understand how the respective states correspond to the supergravity solutions derived in the previous section. We start by looking at the spectrum of non-relativistic string theory and derive the respective worldline descriptions of winding and momentum modes as seen from the nine-dimensional perspective. This is achieved by performing a double-dimensional reduction \cite{Duff:1987bx} of the NR non-linear sigma model---keeping either the winding number or the internal momentum in the compact direction fixed. We do not consider the explicit coupling of the worldline actions to the supergravity theory and the corresponding equations of motion. Instead, we use longitudinal T-duality to conjecture the source terms for the respective DSNC${}^-$ solutions.

\subsection{Non-Relativistic String Theory}

\noindent For the non-relativistic string in a flat target space-time, the spectrum is determined by the following conditions \cite{Gomis:2000bd,Danielsson:2000gi}
\begin{align}
  \label{eq:specNRstring}
      &2\,w\,E = \frac{\ell_s^2}{R}\,k^{A'} k_{A'}\,,&& n\,w=0\,,
\end{align}
if we restrict to the lowest lying non-tachyonic states with $N_L=N_R$ for simplicity. Here, $E$ denotes the energy of the modes and $k^{A'}$ their momentum in the transverse directions. For concrete (super)string expressions see \cite{Danielsson:2000gi, Gomis:2000bd,Kim:2007pc}. The integers $n$ and $w$ correspond to the discrete momentum and winding number along the longitudinal spatial direction that is compactified with radius $R$. It is convenient to define the following charges
\begin{align}\label{eq:bigNbigW}
    &N = \frac{n}{R}&&\mathrm{and}&& W = \frac{w\,R}{\ell_s^2}\,,
\end{align}
so that the dispersion relation for the lowest lying, non-tachyonic state reads $2WE = k_{A'}^2$. The second equation of \eqref{eq:specNRstring} is the level-matching condition. For $w>0$, the first equation corresponds to the mass-shell condition for on-shell states with a non-relativistic dispersion relation and mass $W$. For $w=0$, one can no longer interpret the first equation as an on-shell condition and instead finds $k_{A'}=0$. The so-called ``unwound'' states do, however, play an important role in non-relativistic string theory \cite{Gomis:2000bd,Danielsson:2000gi}, as they correspond to off-shell states, whose propagators behave as $(k^{A'} k_{A'})^{-1}$ for the lowest lying states. These off-shell unwound states thus mediate instantaneous, non-relativistic gravitational interactions.

Let us now consider the sigma model description of the aforementioned modes and their double-dimensional reductions. To do so, we recall the non-relativistic non-linear sigma model \cite{Gomis:2000bd,Bergshoeff:2018yvt}. Here, we focus on the classical part and ignore the dilaton coupling for the moment. The coupling to the background torsional string Newton-Cartan geometry takes the following form
\begin{align}
    S = \frac{1}{4\pi\ell_s^2}\,\int\,d^2\sigma\,\sqrt{|h|}\Big\{&2\,\partial X^\mu\bar\partial X^\nu \, e_{\mu}{}^{A'}e_\nu{}^{B'}\delta_{A'B'} +  \sqrt2\,\bar\lambda(\partial X^\mu)\tau_\mu{}^- + \sqrt2\,\lambda(\bar\partial X^\mu)\tau_\mu{}^+\Big\} \notag\\
     & - 2\,\partial_\sigma X^\mu\partial_\tau X^\nu\,b_{\mu\nu}\,,
\end{align}
where $h_{\alpha\beta}$ denotes the independent worldsheet metric and $h$ its associated determinant. It is furthermore useful to introduce Zweibeine in a lightcone basis $e_\alpha/\bar e_\alpha$ so that $h_{\alpha\beta}=2\,e_{(\alpha}\bar e_{\beta)}$. The inverses are defined as $e^\alpha e_\alpha = 1 = \bar e^\alpha\bar e_\alpha$ and $\bar e^\alpha e_\alpha = 0 = \bar e^\alpha e_\alpha$. These are used to define derivatives $\partial = e^\alpha\partial_\alpha$ and $\bar\partial = \bar e^\alpha\partial_\alpha$.

Let us now assume that there is a longitudinal spatial isometry $k=\partial_z$ in the target space with $z\sim z+2\pi R$. This implies two things: firstly, it allows the string to wrap around this compact direction. To make that more concrete we split the embedding coordinates as $X^\mu(\tau,\sigma) = (X^i,Z)$ and impose
\begin{align}
    &\partial_\sigma X^i =\big(X^i\big)'=0\,, && Z(\sigma+2\pi) = Z(\sigma) + 2\pi R\cdot w\,,
\end{align}
where $w\in\mathds{Z}$. The second condition allows for a linear dependence of the $Z$--field on the spatial coordinate $\sigma\sim\sigma+ 2\pi$ on the closed string worldsheet: $Z' = wR$. This, in turn, allows for defining a winding number $W\sim \int d\sigma Z'$.

Secondly, the existence of a longitudinal isometry affects the structure of the background geometry as shown in section \ref{sec:longitudinalTD}. In particular, the isometry condition implies that all the background fields are independent of the compact embedding coordinate $Z$. It is then natural to decompose the ten-dimensional fields in terms of nine-dimensional torsional Newton-Cartan fields. Using the notation of the reduction Ansatz \eqref{eq:longvielbeinansatz}, we write the relevant pullbacks to the worldline $\dot X^i = \partial_\tau X^i$ as follows:
\begin{align}
    & e= e_{ij}\,\dot X^i\dot X^j\,, && \tau =\tau_i \dot X^i\,, && n = n_i\dot X^i\,,&& m = m_i\dot X^i\,,
\end{align}
where $e_{ij}=\delta_{A'B'}\,e_i{}^{A'}e_j{}^{B'}$. With this, the double dimensional reduction of the classical non-relativistic string theory sigma model over a longitudinal isometry can be written as
\begin{align}\label{eq:GODDR}
   S = \frac{1}{2\ell_s^2}\,\int\,d\tau\,&\kappa\,\Big\{e+\frac12\,\big(\mu+\bar\mu\big)\,\tau+\frac12\,\big(\mu-\bar\mu\big)\big(\dot Z - n\big)\Big\} +Z'\,\big\{\nu - 2\,m\big\}\,,
\end{align}
where we have redefined the following the Lagrange multipliers
\begin{align}
   &\kappa = 2\,e^\tau \bar e^\tau\,\sqrt{|h|}\,,&&\mu = \lambda/ e^\tau\,, &&\bar\mu = \bar\lambda/\bar e^\tau\,,&& \nu=\sqrt{|h|}({\mu\,e^\tau\bar e^\sigma - \bar\mu\,e^\sigma\bar e^\tau})\,.
\end{align}
We assume that all of these are independent of the spatial worldsheet coordinate $\sigma$.
The double dimensional reduction of the non-relativistic string theory \eqref{eq:GODDR} has two independently conserved charges due to the presence of a longitudinal isometry direction. One is the Noether charge of translations in the compact direction, i.e., the discrete internal momentum $N$. The second conserved charge is topological and corresponds to the winding $W$ around the compact direction.
\begin{align}
    &N = \frac{1}{\pi}\,\int\,d\sigma\,\frac{\delta S}{\delta\dot Z} = \frac{\kappa(\mu-\bar\mu)}{2\ell_s^2}\,, && W = \frac{1}{2\pi\ell_s^2}\,\int\,d\sigma\,Z'\,.
\end{align}
The prefactor of the winding charge is chosen so that it agrees with the definition in eq. \eqref{eq:bigNbigW}. The prefactor of the momentum charge is chosen for later convenience. Both charges have units of inverse length.

Let us now consider modes that have either only winding number and zero internal momentum or vice versa. The former are referred to as pure winding modes with $N=0$ and $W\neq 0$. This is equivalent to requiring $\mu=\bar\mu$ and $Z' = \ell_s^2 W$ which implies that $\nu=\mu$ since $e^\tau \bar e^\sigma - e^\sigma\bar e^\tau = 1/\sqrt{|h|}$. Taking everything into account, it leads to the following sigma model\footnote{We note that it is generally not admissable to insert the solution to an equation of motion into the action---in this case, the condition $\mu=\bar\mu$. For the above, however, it turns out to be okay if considering the full set of equations of motion. The ``missing'' equation is effectively eliminating $\dot Z$ in terms of other variables. Compare the comments in section 5 of \cite{Bergshoeff:1996tu}.\label{reinsertcharge}}
\begin{align}
    S = \frac{1}{2\ell_s^2}\,\int\,d\tau\,\Big\{\kappa\,e + \chi\,\big(\tau + \kappa^{-1}Z'\big) - 2\,m\,Z'\Big\}\,,
\end{align}
where $\chi = \kappa\,\mu$. The spectrum formula \eqref{eq:specNRstring} suggests that this model describes massive Bargmann particles in nine dimensions. This is not yet apparent in the present formulation. However, one can integrate out $\kappa$ and $\chi$ (in that order), and finds the following, more recognizable sigma model
\begin{align}\label{eq:GOwinding}
    S = -\frac{W}{2}\int\,d\tau\,\Big\{\frac{e_{ij}\dot X^i\dot X^j}{\tau_i\dot X^i} + 2\,m_i\dot X^i\Big\}\,,
\end{align}
which is the well-known worldline diffeomorphism invariant description of a massive Bargmann particle in a Newton-Cartan background. The gravitational interaction with the Newton potential is captured by the Wess-Zumino term. We note that the mass of the particle is $M_W = W$.

In the spectrum analysis above, we noted that there are also unwound strings that do not appear as asymptotic states. We can nevertheless consider their worldline description by setting $Z'=0$ in \eqref{eq:GODDR}. In addition to that we fix the internal momentum by choosing $\kappa(\mu-\bar\mu) = N/\pi\ell_s$ to a constant value. Using this, we find that the double dimensional reduction of the non-relativistic model \eqref{eq:GODDR} simplifies as follows
\begin{align}\label{eq:masslessGP}
    S = \frac{1}{2\ell_s^2}\,\int\,d\tau\,\Big\{\kappa\,e_{ij}\dot X^i\dot X^j + \rho\,\tau_i\dot X^i - \ell_s^2\,N\,n_i\dot X^i\Big\}\,,
\end{align}
where $\rho = \kappa(\mu+\bar\mu)/2$. We see that it couples to the Kaluza-Klein gauge field $n_i$ via a Wess-Zumino term and charge $Q_N = -N/2$.
For a flat space configuration, given by $e_i{}^{A'}=\delta_i{}^{A'}$, $\tau_i = \delta_i{}^0$ and $n_i=0$ the above reduces to
\begin{align}
    S = \frac{1}{2\ell_s^2}\,\int\,d\tau\,\Big\{\kappa\,\dot X_{A'}^2 + \rho\,\dot X^0\Big)\,.
\end{align}
This action was shown to arise as the non-relativistic limit of a nine-dimensional massless particle \cite{Batlle:2017cfa}.\,\footnote{The non-relativistic limit of a nine-dimensional {\it massive} particle leads to a massless Galilean particle of non-zero colour.} In fact, it was first studied in \cite{souriau1970structure} as a massless Galilean particle with zero colour and zero spin. For simplicity, we will henceforth refer to the sigma model \eqref{eq:masslessGP} as a massless Galilean particle without mentioning that by this we mean a particle of zero colour and spin.

\subsection{Discrete Lightcone Quantization}

\noindent Let us now contrast the spectrum of non-relativistic string theory with that of the discrete light cone quantization at finite internal momentum, see e.g. \cite{Bigatti:1997jy,Seiberg:1997ad,Susskind:1997cw,Hellerman:1997yu}. The basic observation of DLCQ is that string theory with a compact lightlike direction $x^+\sim x^+ + 2\pi R$ and a fixed associated momentum
\begin{align}
    P_+ = \frac{n}{R}\,,
\end{align}
organizes the spectrum in Galilean blocks, where the charge $P_+$ corresponds to the conservation of particle number in non-relativistic systems. This can most easily be seen by considering the mass-shell condition in light-cone coordinates $M^2 = -2\,P_+P_- + k_{A'}^2$.\footnote{For a more careful analysis defining the lightlike compactification as a limit of a spatial one, see \cite{Seiberg:1997ad,Bilal:1998vq}. For a derivation of the longitudinal T-dualities as a limit of the spatial ones, see \cite{Bergshoeff:2019pij}. For an extension of the same procedure to include supersymmetry, see \cite{Lahnsteiner2022}.} When identifying the second lightcone coordinate $x^-$ as the time coordinate and identifying the associated charge as the energy of the system we find that
\begin{align}\label{eq:DLCQspectrum}
    &2\,n\,E = R\,k^{A'}k_{A'}\,, && n\,w=0\,,
\end{align}
if we again restrict to the lowest lying non-tachyonic state with $N_L=N_R$. As above we want to analyze the spectrum for two distinct cases: $n > 0$ and $n = 0$. The first case shows that $E=R\,k_{A'}^2/2n$ and $w=0$ which corresponds to a Bargmann particle with mass $\widetilde N=n/R$. The other case has $n=0$ and thus $k_{A'}=0$. The winding number is unconstrained $w \neq 0$. These null winding states are less intuitive and are usually ignored in a treatment of DLCQ. In \cite{Susskind:1997cw} it was shown that these null winding states are decoupling in the limit of large $n$, but they can still appear as virtual states in scattering processes. In this sense, the states are very much analogous to the unwound string states in non-relativistic string theory. This is more than a coincidence and indeed a consequence of the longitudinal T-duality-like relation between the DLCQ of relativistic string theory and non-relativistic string theory. The duality between the worldsheet theories was first pointed out by \cite{Gomis:2000bd}. This is evidenced by the fact that the spectrum formulas \eqref{eq:DLCQspectrum} and \eqref{eq:specNRstring} are exchanged under
\begin{align}\label{eq:TdualityRtoR}
    &R\quad\leftrightarrow\quad\frac{\ell_s^2}{R}\,, && n\quad\leftrightarrow\quad w\,.
\end{align}
In terms of the rescaled momentum and winding numbers \eqref{eq:bigNbigW} the T-duality rule turns into $N\leftrightarrow W$.

In order to make the above more concrete we consider the discrete light-cone quantization at the level of the sigma model. That is, we start with the relativistic non-linear sigma model on a background with a null isometry.\footnote{The null reduction we consider here should be distinguished from the null reduction of the same type of sigma model considered in \cite{Harmark:2017rpg,Harmark:2019upf}. Here, we consider the sigma model describing specific modes in the spectrum as a source term and reduce the sigma model to a particle action, wheres in \cite{Harmark:2017rpg,Harmark:2019upf} the null reduction was applied in a way as to obtain a non-relativistic string sigma model in one dimension lower.} As shown in section \ref{sec:longitudinalTD}, the NS fields are conveniently decomposed in a basis of Newton-Cartan variables---see in particular, expression \eqref{eq:vielbeinansatz}. Furthermore, we split the embedding coordinates as $X^\mu = (X^i,Z)$ and assume that none of the worldsheet fields depends on the spatial coordinate $\sigma$---except for a linear dependence in $Z$. This ensures that we can consider modes with a non-zero winding number $\int d\sigma Z'$ around the lightlike circle. Denoting $\partial_\tau X^i=\dot X^i$ we can write the pullbacks of the geometric Newton-Cartan type fields as
\begin{align}
    & \tilde e= \delta_{A'B'}\,\tilde e_i{}^{A'}\tilde e_j{}^{B'}\,\dot X^i\dot X^j\,, && \tilde \tau =\tilde \tau_i \dot X^i\,, && \tilde n = \tilde n_i\dot X^i\,,&& \tilde m = \tilde m_i\dot X^i\,.
\end{align}
Using these and the independent worldsheet metric components $\tilde\kappa = \sqrt{-h}h^{\tau\tau}$ and $\tilde\mu = \sqrt{-h}h^{\tau\sigma}$ we can then express the Polyakov model as follows
\begin{align}\label{eq:DLCQparent}
    S = \frac{1}{2\ell_s^2}\,\int\,d\tau\,\tilde\kappa\big\{\tilde e-2\,\tilde \tau\,(\dot Z - \tilde m)\big\} + Z'\,\big\{\tilde\mu\,\tilde \tau - \tilde n\big\}\,.
\end{align}
For $Z'=0$ the above can be recognized to be the action of a massless particle in ten dimensions since the expression in brackets is just the pullback of the ten-dimensional metric to the worldline (in an adapted basis \eqref{eq:vielbeinansatz}). This is consistent with the fact that the source term of a ten-dimensional pp-wave solution is known to be a massless particle.

Due to the presence of a compact direction, we can again define two conserved quantities corresponding to internal momentum and winding, respectively:
\begin{align}
    &\widetilde N = \frac{1}{2\pi}\,\int\,d\sigma\,\frac{\delta S}{\delta \dot Z} = -\frac{\tilde\kappa\tilde\tau}{\ell_s^2}\,,&& \widetilde W = \frac{1}{2\pi\ell_s^2}\int\,d\sigma\,Z'\,.
\end{align}
Since we assumed the background fields, Lagrange multipliers, and $Z'$ to be independent of $\sigma$, we can invert these relations to find $\tilde\kappa = -\ell_s^2\widetilde N/\tilde \tau$ and $Z'= \ell_s^2\widetilde W$. As before, we want to consider pure momentum modes with $\widetilde N\neq 0$/$\widetilde W=0$ and winding modes with $\widetilde N=0$/$\widetilde W\neq 0$ separately. We expect the former to describe nine-dimensional massive particles and the latter to describe massless Galilean particles as anticipated above. Let us start by considering the momentum modes. We eliminate\footnote{See footnote \ref{reinsertcharge}.} the Lagrange multiplier $\tilde \kappa$ in terms of the momentum number and the pullback of the clock one-form to find
\begin{align}
    S = -\frac{\widetilde N}{2}\,\int\,d\tau\,\Big\{\frac{\tilde e_{ij}\dot X^i\dot X^j}{\tilde\tau_i \dot X^i} + 2\,\tilde m_i\dot X^i\Big\}\,.
\end{align}
This is again the sigma model description of a massive Bargmann particle with mass $\tilde M_N = \widetilde N$. It is formally equivalent to the reduction of the non-relativistic winding mode \eqref{eq:GOwinding} upon applying the longitudinal T-duality rules \eqref{identifications}. The masses of both models are related by the action of the T-duality rules \eqref{eq:TdualityRtoR}
\begin{align}
    M_W = W \qquad\leftrightarrow\qquad \tilde M_N = \widetilde N\,,
\end{align}
We remind the reader that $\widetilde N=n/R$ and $W=wR/\ell_s^2$ and thus $W\leftrightarrow \widetilde N$ under the T-duality rule \eqref{eq:TdualityRtoR}. This shows very concretely that the winding modes of non-relativistic string theory and the momentum modes that occur in the DLCQ of relativistic string theory are indeed T-dual to each other.

It is well-known that the pp-wave solution of NS gravity is sourced by a ten-dimensional massless particle. We have seen that this corresponds to the momentum mode in DLCQ. Under longitudinal T-duality, this maps to the winding mode of non-relativistic string theory. At the same time, we showed that the pp-wave solution is related to the winding string solution via longitudinal T-duality. Taken together, these observations provide evidence that the winding string solution \eqref{sec:windingstring} is indeed sourced by the winding mode of non-relativistic string theory. We have not derived the explicit matter-coupled equations of motion to make this statement concrete. One option of doing so systematically would be to adapt the expansion techniques developed in \cite{Hansen:2020pqs} to DSNC${}^-$ gravity. It would be interesting to return to the question of matter coupling in in the future.

If we, instead, consider the DLCQ of the Polyakov model \eqref{eq:DLCQparent} at zero internal momentum $\widetilde N=0$, we find the following non-linear sigma model
\begin{align}
    S = \frac{1}{2\ell_s^2}\,\int\,d\tau\,\Big\{\tilde\kappa\,\tilde e_{ij}\dot X^i\dot X^j + \tilde \rho\,\tilde\tau_i\dot X^i - \ell_s^2\widetilde W\,\tilde n_i\dot X^i\Big\}\,,
\end{align}
where $\tilde \rho = \tilde\mu\,Z'$. Again, the answer is a massless Galilean particle coupled to the $\tilde n_i$ field with charge $\widetilde Q_W = -\widetilde W/2$. This answer is formally equivalent to the reduction of the momentum mode of the non-relativistic theory and thus in agreement with longitudinal T-duality \eqref{eq:TdualityRtoR}. Here, the charges are exchanged
\begin{align}
  Q_N = -\frac{N}{2}\qquad\leftrightarrow\qquad \widetilde{Q}_{\widetilde W} = -\frac{\widetilde W}{2}
\end{align}
upon exchanging internal momentum with winding $N\leftrightarrow \widetilde W$. In alignment with our previous comments, we see this as the source of the fundamental string \eqref{RF1} and unwound string solution \eqref{eq:unwoundstring} respectively. We have seen that the supergravity solution has intrinsic torsion---when seen as a solution of DSNC${}^-$ or torsional Newton-Cartan backgrounds. It would be interesting to see whether this fact can be derived directly from the worldline description of the source.

Let us now provide evidence why we consider the massless Galilei particle to be the source for the unwound string solution \eqref{eq:unwoundstring}. We use the fact that the relativistic fundamental string solution \eqref{RF1} is sourced by the fundamental string sigma model itself. In the presence of a compact direction this corresponds to the winding mode in DLCQ. Using longitudinal T-duality we thus conjecture that the unwound string solution \eqref{eq:unwoundstring} is sourced by the momentum mode of non-relativistic string theory.

This concludes our discussion of longitudinal T-duality at the level of the non-linear sigma model description. We have seen that winding modes of non-relativistic string theory and momentum modes of DLCQ string theory get exchanged under longitudinal T-duality. Both reduce to Galilean particles of mass $M_W=\widetilde M_{\widetilde N}$ in nine dimensions. The other modes---momentum in non-relativistic string theory and winding in DLCQ---do not appear as asymptotic states in scattering amplitudes. These seemingly unphysical modes do, however, appear as intermediate states giving a physical interpretation to the instantaneous Newtonian gravitational interaction \cite{Danielsson:2000mu} between asymptotic states. As such, they also have a sigma model description which is known as massless Galilean particles of zero color, and zero spin \cite{Batlle:2017cfa}. They couple to the $\mathsf{U}(1)-$gauge field $n_i=\tilde n_i$, which is not part of the nine-dimensional Newton-Cartan geometry, with charge $Q_N=\widetilde Q_{\widetilde W}$. The coupling to the respective Kaluza-Klein and Kalb-Ramond vectors suggests that the massive and massless Galilean particle act as sources for the supergravity solutions derived in the previous section as follows
\begin{align}
    \mathrm{Massive~Bargmann~Particle~\eqref{eq:GOwinding}}\quad&\leftrightsquigarrow\quad \mathrm{Winding~String~Solution~\ref{sec:windingstring}}\,,\notag\\
    \mathrm{Massless~Galilean~Particle~\eqref{eq:masslessGP}}\quad&\leftrightsquigarrow\quad \mathrm{Unwound~String~Solution~\ref{sec:unwoundstring}}\,.\notag
\end{align}
We have seen that lightlike T-duality exchanges the winding and unwound string solutions of DSNC${}^-$ supergravity. Given the relation to the worldline actions for the winding and momentum modes of non-relativistic string theory, it is natural to ask whether this duality can be seen directly at the level of either the spectrum formulas \eqref{eq:specNRstring}/\eqref{eq:DLCQspectrum}---or, at the level of the worldline sigma models \eqref{eq:GOwinding}/\eqref{eq:masslessGP}.

\section{Conclusions}\label{sec:conclusions}

\noindent In this paper, we have investigated the basic half-supersymmetric NS brane solutions of non-relativistic (super)string theory using longitudinal T-duality as a solution generating transformation. Originally, in the relativistic case T-duality was only considered with respect to a spatial direction leading to the well-known Buscher rules. Later, T-duality with respect to a timelike direction was considered  leading to supergravity theories with a different signature \cite{Hull:1998vg}. In this paper, we have considered the third option, namely T-duality with respect to a lightlike direction. The requirement that the isometry is lightlike truncates the relativistic supergravity theory to a non-relativistic one, which we have denominated {\sl $\mathcal{N} = (1,0)_0$}  supergravity and satifies the constraint \eqref{eq:(1,0)constraints}. The bosonic part of the corresponding non-relativistic T-duality rules were derived in \cite{Bergshoeff:2018yvt}.

To find the basic half-supersymmetric brane solutions of DSNC${}^-$ supergravity, we have made use of the  observation that each solution of the relativistic $\mathcal{N} = (1,0)$ supergravity theory that preserves at least one supercharge has a lightlike isometry \cite{Fontanella:2019avn}.  This implies that each relativistic supersymmetric solution
that satisfies the geometric constraint $Z_{ij}=0$, see eq.~\ref{eq:(1,0)constraints},  is also a supersymmetric solution of the non-relativistic $\mathcal{N} = (1,0)_0$  supergravity theory.
Using the fact that the 4 basic relativistic half-supersymmetric NS brane solutions (fundamental string, pp-wave, NS 5-brane and KK5 monopole) all satisfy the $\tilde Z_{ij}=0$ constraint, we  found the 4 basic half-supersymmetric NS brane solutions of the DSNC${}^-$ supergravity theory, by applying longitudinal T-duality.

Focussing on the perturbative string solutions, we found a half-supersymmetric winding string and a half-supersymmetric unwound string solution that have a rather different interpretation. The winding string is part of the on-shell spectrum of non-relativistic string theory. Its longitudinal T-dual version is the pp-wave and its nine-dimensional version is a massive Bargmann particle. On the other hand the unwound string only makes sense off-shell where it mediates the instantaneous gravitational force. It is the longitudinal T-dual of a winding fundamental string solution of NS${}_0$ gravity  while its nine-dimensional version is a massless Galilean particle.

The fact that we have two different 0-branes in nine dimensions, namely a Bargmann particle and a  Galilean particle, implies that there is no $\mathsf{O}(1,1)$ T-duality transformation that transforms these two 0-branes into each other, like it happens for  relativistic $\mathcal{N}=1$ T-duality. This is consistent with the fact that the longitudinal T-duality maps one theory to another theory instead of mapping it onto itself. To obtain T-duality symmetries like in the relativistic case, one could  consider a lightlike T-duality transformation which maps DSNC${}^-$ gravity onto itself. This is consistent with the fact that under a lightlike T-duality the winding string and unwound string solutions are transformed into each other.

Following \cite{Blair:2021waq,Ebert:2021mfu}, it is interesting to speculate about the extension of our results to non-relativistic IIA/B and heterotic superstring theory and to consider NR D$p$-brane solutions. It seems that these solutions can occur both as non-perturbative soliton-like solutions of NR IIA and IIB supergravity with a co-dimension two foliation or as fundamental brane solutions of new NR supergravity theories with a co-dimension $p+1$ foliation. We will denote these different  NR supergravity theories as (IIA,$p+1$) supergravity for $p$ even and (IIB, $p+1$) supergravity for $p$ odd.

Concerning the  realization of NR Dp-branes as soliton-like solutions, one would expect that the NR D1-brane solution can be obtained from a NR S-duality of the winding string solution we constructed in this work embedded into (IIB,2)-supergravity.\,\footnote{Alternatively, one could obtain this NR D1-brane solution starting from a D2-brane solution of IIA${}_0$ supergravity, which is an extension of NS${}_0$ supergravity  to the IIA case, and applying a longitudinal T-duality transformation.} Next, the other NR Dp-brane solutions can be obtained, as solutions of (IIA,2) or (IIB,2) supergravity for p even or odd, respectively, by applying an extension of NR  transverse T-duality to the RR potentials \cite{Ebert:2021mfu}.

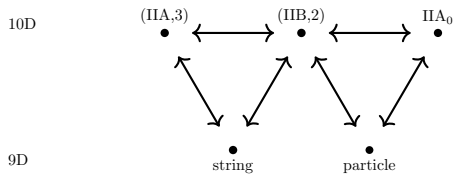
\begin{figure}[h]
 \centering
\hskip 0truecm
  \begin{tikzpicture}[thick,scale=0.6, every node/.style={scale=0.6}]
       \node[draw,circle,right = 6 cm=l, fill=black,scale=0.2,label=above:{(IIA,3)}] (c) {C};
          \node[draw,circle,right = 9 cm=m, fill=black,scale=0.2,label=above:{(IIB,2)}] (d) {D};
           \node[draw,circle,right = 12 cm=m, fill=black,scale=0.2,label=above:{ IIA${}_0$}] (e)  {E};

\node[draw,circle,below =2.59cm,right=7.5 cm, fill=black, scale=0.2,label=below:{string}] (h) {H};
\node[draw,circle,below =2.59cm,right=10.5 cm, fill=black, scale=0.2,label=below:{particle}] (j) {J};

%

\draw[<->,shorten <=9pt,shorten >=9pt] (c) to (h);
\draw[<->,shorten <=9pt,shorten >=9pt] (d) to (h);\draw[<->,shorten <=9pt,shorten >=9pt] (c) to (d);

\draw[<->,shorten <=9pt,shorten >=9pt] (d) to (j);
\draw[<->,shorten <=9pt,shorten >=9pt] (e) to (j);\draw[<->,shorten <=9pt,shorten >=9pt] (d) to (e);

\filldraw (+2.5,0.2) node[anchor=west] {10D};
\filldraw (+2.5,-2.8) node[anchor=west] {9D};

  \end{tikzpicture}
\caption{this Figure pictures the membrane T-duality discussed in the text. The longitudinal T-duality discussed in this paper is represented by the right triangle while the  left triangle represents the membrane T-duality.  This membrane T-duality  maps (solutions of) (IIA,3) supergravity to (IIB,2) supergravity. }
\label{scan}
\end{figure}

We expect that the NR D2-brane  not only occurs as a non-perturbative solution of (IIA,2) supergravity but also as a fundamental solution of a new (IIA,3) supergravity theory. This new supergravity theory can be obtained from a transversal dimensional reduction of a membrane-limit of M-theory \cite{Blair:2021waq}. Assuming that the double dimensional reduction of (IIA,3) supergravity gives the same NR nine-dimensional supergravity theory as the direct dimensional reduction of (IIB,2) supergravity, this would establish a {\it membrane T-duality} mapping (solutions of)  (IIB,2) supergravity to  (IIA,3) supergravity, see Figure \ref{scan}.  We hope to make some of these speculations  concrete in a forthcoming work.
\vskip .3truecm

\section*{Acknowledgements}

\noindent We thank Kevin Grosvenor  for valuable discussions and Ziqi Yan for useful comments. JL would like to thank Gerben Oling, Niels Obers, and Ziqi Yan for inspiring discussions at NORDITA. Beyond that, we especially thank Quim Gomis and Tonnis ter Veldhuis, who were involved in the initial stage of this project. Above all, however, we would like to express our gratitude to Ceyda \c{S}im\c{s}ek, who has made numerous significant contributions to this project. The work of LR has been initially supported by the FOM/NWO free program Scanning New Horizons and successively supported by Next Generation EU through the Maria Zambrano grant from the Spanish Ministry of Universities under the Plan de Recuperacion, Transformacion y Resiliencia.

\appendix

\section{Conventions} \label{sec:conventions}
\noindent In this work, we are following the conventions of \cite{Bergshoeff:2021tfn}. We denote the ten-dimensional coordinates by $x^\mu$. We will consider two kinds of geometries: semi-Riemannian geometries with NS fields and non-Lorentzian geometries with a co-dimension two foliation denoted by DSNC${}^-$, see also the remarks in the introduction and appendix \ref{sec:TSNC}. In both cases, we are considering isometries that are either spatial or lightlike. For any practical calculation, it is then useful to split and adapt the coordinates such that the Killing vector takes a simple form. Here is an overview of the different choices:
\begin{align}
  & \mathrm{Geometry} &&\mathrm{Coordinates} &&\mathrm{Killing~Vector} && \mathrm{Type}\notag\\
  & NS && x^\mu = (y,x^i)= (y,x^0,z_{(8)}^m)\,, && k^\mu\partial_\mu = \partial_y\,, && G(k,k)>0\,,\notag\\
  & NS && x^\mu = (z,x^i)= (z, u,z_{(8)}^m)\,, && k^\mu\partial_\mu = \partial_z\,,  && G(k,k) = 0\,,\notag\\
  & DSNC{}^-  && x^\mu = (y,x^i)= (z, t,z_{(8)}^m)\,, && k^\mu\partial_\mu = \partial_z\,,  && \tau(k,k)>0\,,\notag\\
  & DSNC{}^-   && x^\mu = (y,x^i)= (y,x^0,z_{(8)}^m)\,,&& k^\mu\partial_\mu = \partial_y\,,  && e(k,k)>0\,,\notag\\
  & DSNC{}^-   && x^\mu = (u,x^i)= (u,v,z_{(8)}^m)\,,  && \ell^\mu\partial_\mu = \partial_u\,,&& \tau(\ell,\ell)=0\,,
\end{align}
where we have defined null directions for example as $\partial_y = 2^{-1/2}(\partial_t + \partial_z)$, etc. In the text we sometimes use the following notation for the flat transversal metric
\begin{align}
    dz_{(8)}^2 = (dz^2)^2 + (dz^3)^2 + \cdots + (dz^9)^2\,.
\end{align}
Ten-dimensional flat Lorentz indices are denoted by $\hat A$ and split into $(A,A')$, where $A=0,1$ and $A'=2,\cdots,9$, unless specified otherwise. We refer to this as a splitting into longitudinal $A$ and transversal $A'$ directions. We use the `mostly plus' form of the Minkowski metric, i.e., $(\eta_{\hat A\hat B}) = (-++\cdots +)$. Symmetrization $(AB)$ and antisymmetrization $[AB]$ are defined with weight one, and we denote the symmetric traceless combinations with curly brackets $S_{\{AB\}} = S_{(AB)} - 1/2\,\eta_{AB}S_C{}^C$.

\section{Torsional String Newton-Cartan Geometry} \label{sec:TSNC}

\noindent In this section, we give some details on the non-Lorentzian geometric structures appearing in this paper. We will use the name torsional string Newton-Cartan for generic geometric structures without any further geometric constraints on the torsion. The self-dual DSNC${}^-$ geometry that is relevant in this paper is a special case where the torsion tensor satisfies $T_{\mu\nu}^\rho\,\tau_\rho{}^-=0$. We refer the reader to appendices B and C of \cite{Bergshoeff:2021bmc} for more details. See also \cite{Bidussi:2021ujm} and references therein.

The main feature that distinguishes string Newton-Cartan structures from other geometric strucures is the fact that the Kalb-Ramond field strength $h_{\mu\nu\rho}=3\,\partial_{[\mu}b_{\nu\rho]}$ and that of the dilaton $\partial_\mu\phi$ do not transform covariantly under the local symmetries \eqref{eq:symmsDSNCm}. Instead they transform with a derivative of the boost and an-isotropic dilatation parameter, respectively. In other words, they are part of the dependent spin connections
\begin{subequations}\label{eq:galspinconn}
\begin{align}
b_\mu &= e_\mu{}^{A'}\,\tau_{A'A}{}^A +\tau_\mu{}^A\partial_A\phi \,,\\
\omega_\mu &= \big(\,\tau_\mu{}^{AB}-\frac12\,\tau_\mu{}^C\tau^{AB}{}_C \big)\epsilon_{AB} - \tau_\mu{}^A\,\epsilon_{AB}\partial^B\phi\,,\\
\omega_\mu{}^{AA'} &= -e_\mu{}^{AA'}+e_{\mu B'}e^{AA'B'} + \frac12\,\epsilon^A{}_B\,h_\mu{}^{BA'}  + \tau_{\mu B} W^{BAA'} \,,\\
\omega_\mu{}^{A'B'} &= -2\, e_{\mu}{}^{[A'B']}+e_{\mu C'}e^{A'B'C'} - \frac12\,\tau_\mu{}^A\,\epsilon_{AB}\,h^{BA'B'} \,,
\end{align}
\end{subequations}
where $\tau_{\mu\nu}{}^A=\partial_{[\mu}\tau_{\nu]}{}^A$ and $e_{\mu\nu}{}^{A'} = \partial_{[\mu} e_{\nu]}{}^{A'}$. We note that not all components can be solved for, which is reflected by the undetermined $W^{ABA'}$ component which is traceless symmetric in the $(A,B)$ indices, but otherwise arbitrary. Since all the relevant expressions---such as action, equations of motion, symmetry transformation rules, and Killing spinor equations---follow from a limit it is clear that nothing depends on $W$, see also \cite{Bergshoeff:2021bmc}. We can furthermore introduce connections $\Gamma_{\mu\nu}^\rho$ by imposing Vielbein postulates
\begin{subequations}\label{eq:metcomp}
\begin{align}
\nabla_\mu\tau_\nu{}^A &\equiv \partial_\mu\tau_\nu{}^A - \omega_\mu\,\epsilon^{AB}\tau_{\nu B} - b_\mu\,\tau_\nu{}^A - \Gamma_{\mu\nu}^\rho\tau_\rho{}^A = 0\,,\label{eq:taucomp}\\
\nabla_\mu e_\nu{}^{A'} &\equiv \partial_\mu e_\nu{}^{A'} - \omega_\mu{}^{A'B'} e_{\nu B'} + \omega_\mu{}^{AA'}\tau_{\nu A} - \Gamma_{\mu\nu}^\rho e_\rho{}^{A'}=0\,,\label{eq:ecomp}
\end{align}
\end{subequations}
The symbols $\Gamma_{\mu\nu}^\rho$ are useful expressions and allow us to define diffeomorphism covariant derivatives. However, there is one caveat worth pointing out: due to the non-trivial symmetry structure, these connections are not invariant under local boosts \cite{Bergshoeff:2021tfn}. Hence we do not refer to $\Gamma_{\mu\nu}^\rho$ as affine connections. However, it turns out that the antisymmetric part is invariant and thus makes sense as a torsion tensor $[\nabla_\mu,\nabla_\nu]f = T_{\mu\nu}^\rho\nabla_\rho f$. By considering the antisymmetric part of \eqref{eq:metcomp} it is not hard to see that the torsion is non-zero even without coupling to matter. This implies that generic torsional string Newton-Cartan geometries have intrinsic torsion \cite{Figueroa-OFarrill:2020gpr}
\begin{align}
T_{\mu\nu}^\rho &= 2\,\Gamma_{[\mu\nu]}^\rho = \rmR_{\mu\nu}(H^A)\tau_A{}^\rho\,,\notag\\
\mathrm{with}\qquad \rmR_{\mu\nu}(H^A) &  = e_\mu{}^{A'}e_\nu{}^{B'}\,\tau_{A'B'}{}^A + 2\,e_{[\mu}{}^{A'}\tau_{\nu]B}\,\tau_{A'}{}^{\{BA\}}\,,\label{eq:torsion2Form}
\end{align}
where $\rmR_{\mu\nu}(H^A)\equiv 2\,\partial_{[\mu}\tau_{\nu]}{}^A - 2\,\big(\epsilon^A{}_B\,\omega_{[\mu} +\delta^A{}_{B}\,b_{[\mu}\big)\tau_{\nu]}{}^B$ is the covariant version of $\tau_{\mu\nu}{}^A$. With this we can define DSNC${}^-$ geometries as special torsional string Newton-Cartan structures with half of the intrinsic torsion set to zero
\begin{align}
      &T_{\mu\nu}^\rho\,\tau_\rho{}^- = 0\,, &&\leftrightarrow&& \rmR_{\mu\nu}(H^-)=0\,, &&\leftrightarrow&& \tau_{[\mu}{}^-\partial_\nu^{}\tau_{\rho]}{}^-=0\,,
\end{align}
as anticipated in the introduction \eqref{constraint2}. We have shown in \cite{Bergshoeff:2021tfn} and reviewed in section \ref{sec:susyTduality} that these constraints are special in the sense that they are the maximal set of constraints consistent with superymmetry. More precisely, we have seen that they are actually a necessary requirement for the consistency of the multiplet.

Apart from the ``metric-compatible'' covariant derivative $\nabla_\mu$ as defined in \eqref{eq:taucomp} and \eqref{eq:ecomp} we use one other covariant derviative $\mathcal D_\mu$. It is covariant with respect to $\mathsf{SO}(1,1)\times \mathsf{SO}(8)$ and dilatations, but not Galilean boosts. To clarify the difference, let us write \eqref{eq:ecomp} in the following equivalent ways: $\nabla_\mu e_\nu{}^{A'} = \mathcal D_\mu e_\nu{}^{A'} + \omega_\mu{}^{AA'}\tau_{\nu A}- \Gamma_{\mu\nu}^\rho e_\rho{}^{A'}$.


\begin{thebibliography}{10}

\bibitem{Gomis:2000bd}
J.~Gomis and H.~Ooguri, \emph{{Nonrelativistic closed string theory}},
  \href{https://doi.org/10.1063/1.1372697}{\emph{J.\ Math.\ Phys.} {\bfseries
  42} (2001) 3127} [\href{https://arxiv.org/abs/hep-th/0009181}{{\ttfamily
  hep-th/0009181}}].

\bibitem{Danielsson:2000gi}
U.~H. Danielsson, A.~Guijosa and M.~Kruczenski, \emph{{IIA/B, wound and
  wrapped}}, \href{https://doi.org/10.1088/1126-6708/2000/10/020}{\emph{JHEP}
  {\bfseries 10} (2000) 020}
  [\href{https://arxiv.org/abs/hep-th/0009182}{{\ttfamily hep-th/0009182}}].

\bibitem{Gomis:2005pg}
J.~Gomis, J.~Gomis and K.~Kamimura, \emph{{Non-relativistic superstrings: A New
  soluble sector of $AdS_5 \times S^5$}},
  \href{https://doi.org/10.1088/1126-6708/2005/12/024}{\emph{JHEP} {\bfseries
  12} (2005) 024} [\href{https://arxiv.org/abs/hep-th/0507036}{{\ttfamily
  hep-th/0507036}}].

\bibitem{Bergshoeff:2018yvt}
E.~Bergshoeff, J.~Gomis and Z.~Yan, \emph{{Nonrelativistic String Theory and
  T-Duality}}, \href{https://doi.org/10.1007/JHEP11(2018)133}{\emph{JHEP}
  {\bfseries 11} (2018) 133}
  [\href{https://arxiv.org/abs/1806.06071}{{\ttfamily 1806.06071}}].

\bibitem{Bergshoeff:2019pij}
E.~A. Bergshoeff, J.~Gomis, J.~Rosseel, C.~\c{S}im\c{s}ek and Z.~Yan,
  \emph{{String Theory and String Newton-Cartan Geometry}},
  \href{https://doi.org/10.1088/1751-8121/ab56e9}{\emph{J.\ Phys.\ A}
  {\bfseries 53} (2020) 014001}
  [\href{https://arxiv.org/abs/1907.10668}{{\ttfamily 1907.10668}}].

\bibitem{Bergshoeff:2021bmc}
E.~A. Bergshoeff, J.~Lahnsteiner, L.~Romano, J.~Rosseel and C.~\c{S}im\c{s}ek,
  \emph{{A non-relativistic limit of NS-NS gravity}},
  \href{https://doi.org/10.1007/JHEP06(2021)021}{\emph{JHEP} {\bfseries 06}
  (2021) 021} [\href{https://arxiv.org/abs/2102.06974}{{\ttfamily
  2102.06974}}].

\bibitem{Bidussi:2021ujm}
L.~Bidussi, T.~Harmark, J.~Hartong, N.~A. Obers and G.~Oling, \emph{{Torsional
  string Newton-Cartan geometry for non-relativistic strings}},
  \href{https://doi.org/10.1007/JHEP02(2022)116}{\emph{JHEP} {\bfseries 02}
  (2022) 116} [\href{https://arxiv.org/abs/2107.00642}{{\ttfamily
  2107.00642}}].

\bibitem{Harmark:2017rpg}
T.~Harmark, J.~Hartong and N.~A. Obers, \emph{{Nonrelativistic strings and
  limits of the AdS/CFT correspondence}},
  \href{https://doi.org/10.1103/PhysRevD.96.086019}{\emph{Phys.\ Rev.\ D}
  {\bfseries 96} (2017) 086019}
  [\href{https://arxiv.org/abs/1705.03535}{{\ttfamily 1705.03535}}].

\bibitem{Harmark:2018cdl}
T.~Harmark, J.~Hartong, L.~Menculini, N.~A. Obers and Z.~Yan, \emph{{Strings
  with Non-Relativistic Conformal Symmetry and Limits of the AdS/CFT
  Correspondence}}, \href{https://doi.org/10.1007/JHEP11(2018)190}{\emph{JHEP}
  {\bfseries 11} (2018) 190}
  [\href{https://arxiv.org/abs/1810.05560}{{\ttfamily 1810.05560}}].

\bibitem{Harmark:2019upf}
T.~Harmark, J.~Hartong, L.~Menculini, N.~A. Obers and G.~Oling, \emph{{Relating
  non-relativistic string theories}},
  \href{https://doi.org/10.1007/JHEP11(2019)071}{\emph{JHEP} {\bfseries 11}
  (2019) 071} [\href{https://arxiv.org/abs/1907.01663}{{\ttfamily
  1907.01663}}].

\bibitem{Gomis:2020fui}
J.~Gomis, Z.~Yan and M.~Yu, \emph{{Nonrelativistic Open String and Yang-Mills
  Theory}}, \href{https://doi.org/10.1007/JHEP03(2021)269}{\emph{JHEP}
  {\bfseries 03} (2021) 269}
  [\href{https://arxiv.org/abs/2007.01886}{{\ttfamily 2007.01886}}].

\bibitem{Gomis:2020izd}
J.~Gomis, Z.~Yan and M.~Yu, \emph{{T-Duality in Nonrelativistic Open String
  Theory}}, \href{https://doi.org/10.1007/JHEP02(2021)087}{\emph{JHEP}
  {\bfseries 02} (2021) 087}
  [\href{https://arxiv.org/abs/2008.05493}{{\ttfamily 2008.05493}}].

\bibitem{Kluson:2018egd}
J.~Kluso\v{n}, \emph{{Remark About Non-Relativistic String in Newton-Cartan
  Background and Null Reduction}},
  \href{https://doi.org/10.1007/JHEP05(2018)041}{\emph{JHEP} {\bfseries 05}
  (2018) 041} [\href{https://arxiv.org/abs/1803.07336}{{\ttfamily
  1803.07336}}].

\bibitem{Kluson:2018vfd}
J.~Kluso\v{n}, \emph{{Note About T-duality of Non-Relativistic String}},
  \href{https://doi.org/10.1007/JHEP08(2019)074}{\emph{JHEP} {\bfseries 08}
  (2019) 074} [\href{https://arxiv.org/abs/1811.12658}{{\ttfamily
  1811.12658}}].

\bibitem{Kluson:2019ifd}
J.~Kluso\v{n}, \emph{{$(m,n)$-String and D1-Brane in Stringy Newton-Cartan
  Background}}, \href{https://doi.org/10.1007/JHEP04(2019)163}{\emph{JHEP}
  {\bfseries 04} (2019) 163}
  [\href{https://arxiv.org/abs/1901.11292}{{\ttfamily 1901.11292}}].

\bibitem{Roychowdhury:2019qmp}
D.~Roychowdhury, \emph{{Probing tachyon kinks in Newton-Cartan background}},
  \href{https://doi.org/10.1016/j.physletb.2019.06.031}{\emph{Phys.\ Lett.\ B}
  {\bfseries 795} (2019) 225}
  [\href{https://arxiv.org/abs/1903.05890}{{\ttfamily 1903.05890}}].

\bibitem{Hartong:2021ekg}
J.~Hartong and E.~Have, \emph{{Nonrelativistic Expansion of Closed Bosonic
  Strings}}, \href{https://doi.org/10.1103/PhysRevLett.128.021602}{\emph{Phys.
  Rev. Lett.} {\bfseries 128} (2022) 021602}
  [\href{https://arxiv.org/abs/2107.00023}{{\ttfamily 2107.00023}}].

\bibitem{Christensen:2013lma}
M.~H. Christensen, J.~Hartong, N.~A. Obers and B.~Rollier, \emph{{Torsional
  Newton-Cartan Geometry and Lifshitz Holography}},
  \href{https://doi.org/10.1103/PhysRevD.89.061901}{\emph{Phys. Rev. D}
  {\bfseries 89} (2014) 061901}
  [\href{https://arxiv.org/abs/1311.4794}{{\ttfamily 1311.4794}}].

\bibitem{Bergshoeff:2021xxx}
E.~Bergshoeff, J.~Lahnsteiner, L.~Romano, J.~Rosseel and K.~van Helden,
  \emph{{Torsional String Newton Cartan Geometry}}, {\emph{to appear} }.

\bibitem{Figueroa-OFarrill:2020gpr}
J.~Figueroa-O'Farrill, \emph{{On the intrinsic torsion of spacetime
  structures}},  \href{https://arxiv.org/abs/2009.01948}{{\ttfamily
  2009.01948}}.

\bibitem{Bergshoeff:2021tfn}
E.~A. Bergshoeff, J.~Lahnsteiner, L.~Romano, J.~Rosseel and C.~Simsek,
  \emph{{Non-Relativistic Ten-Dimensional Minimal Supergravity}},
  \href{https://arxiv.org/abs/2107.14636}{{\ttfamily 2107.14636}}.

\bibitem{Gomis:2019zyu}
J.~Gomis, J.~Oh and Z.~Yan, \emph{{Nonrelativistic String Theory in Background
  Fields}}, \href{https://doi.org/10.1007/JHEP10(2019)101}{\emph{JHEP}
  {\bfseries 10} (2019) 101}
  [\href{https://arxiv.org/abs/1905.07315}{{\ttfamily 1905.07315}}].

\bibitem{Yan:2019xsf}
Z.~Yan and M.~Yu, \emph{{Background Field Method for Nonlinear Sigma Models in
  Nonrelativistic String Theory}},
  \href{https://doi.org/10.1007/JHEP03(2020)181}{\emph{JHEP} {\bfseries 03}
  (2020) 181} [\href{https://arxiv.org/abs/1912.03181}{{\ttfamily
  1912.03181}}].

\bibitem{Gallegos:2019icg}
A.~D. Gallegos, U.~G\"ursoy and N.~Zinnato, \emph{{Torsional Newton Cartan
  gravity from non-relativistic strings}},
  \href{https://doi.org/10.1007/JHEP09(2020)172}{\emph{JHEP} {\bfseries 09}
  (2020) 172} [\href{https://arxiv.org/abs/1906.01607}{{\ttfamily
  1906.01607}}].

\bibitem{Polchinski:1998rq}
J.~Polchinski, \emph{{String theory. Vol. 1: An introduction to the bosonic
  string}}, Cambridge Monographs on Mathematical Physics. Cambridge University
  Press, 12, 2007,
  \href{https://doi.org/10.1017/CBO9780511816079}{10.1017/CBO9780511816079}.

\bibitem{Yan:2021hte}
Z.~Yan and M.~Yu, \emph{{KLT Factorization of Nonrelativistic String
  Amplitudes}},  \href{https://arxiv.org/abs/2112.00025}{{\ttfamily
  2112.00025}}.

\bibitem{Klebanov:2000pp}
I.~R. Klebanov and J.~M. Maldacena, \emph{{(1+1)-dimensional NCOS and its U(N)
  gauge theory dual}},
  \href{https://doi.org/10.1142/S0217751X01004001}{\emph{Adv. Theor. Math.
  Phys.} {\bfseries 4} (2000) 283}
  [\href{https://arxiv.org/abs/hep-th/0006085}{{\ttfamily hep-th/0006085}}].

\bibitem{Goroff:1983hc}
M.~Goroff and J.~H. Schwarz, \emph{{$D$-dimensional Gravity in the Light Cone
  Gauge}}, \href{https://doi.org/10.1016/0370-2693(83)91630-1}{\emph{Phys.
  Lett. B} {\bfseries 127} (1983) 61}.

\bibitem{Duval:1984cj}
C.~Duval, G.~Burdet, H.~K{\"u}nzle and M.~Perrin, \emph{{Bargmann Structures
  and Newton-cartan Theory}},
  \href{https://doi.org/10.1103/PhysRevD.31.1841}{\emph{Phys.\ Rev.\ D}
  {\bfseries 31} (1985) 1841}.

\bibitem{Julia:1994bs}
B.~Julia and H.~Nicolai, \emph{{Null Killing vector dimensional reduction and
  Galilean geometrodynamics}},
  \href{https://doi.org/10.1016/0550-3213(94)00584-2}{\emph{Nucl.\ Phys.\ B}
  {\bfseries 439} (1995) 291}
  [\href{https://arxiv.org/abs/hep-th/9412002}{{\ttfamily hep-th/9412002}}].

\bibitem{Bergshoeff:2017dqq}
E.~Bergshoeff, A.~Chatzistavrakidis, L.~Romano and J.~Rosseel,
  \emph{{Newton-Cartan Gravity and Torsion}},
  \href{https://doi.org/10.1007/JHEP10(2017)194}{\emph{JHEP} {\bfseries 10}
  (2017) 194} [\href{https://arxiv.org/abs/1708.05414}{{\ttfamily
  1708.05414}}].

\bibitem{Fontanella:2019avn}
A.~Fontanella and T.~Ort\'\i{}n, \emph{{On the supersymmetric solutions of the
  Heterotic Superstring effective action}},
  \href{https://doi.org/10.1007/JHEP10(2021)130}{\emph{JHEP} {\bfseries 06}
  (2020) 106} [\href{https://arxiv.org/abs/1910.08496}{{\ttfamily
  1910.08496}}].

\bibitem{souriau1970structure}
J.-M. Souriau, \emph{Structure des systemes dynamiques, dunod, paris;
  translation: Structure of dynamical systems}, {\emph{Progress in Mathematics}
  {\bfseries 149} (1970) }.

\bibitem{Batlle:2017cfa}
C.~Batlle, J.~Gomis, L.~Mezincescu and P.~K. Townsend, \emph{{Tachyons in the
  Galilean limit}}, \href{https://doi.org/10.1007/JHEP04(2017)120}{\emph{JHEP}
  {\bfseries 04} (2017) 120}
  [\href{https://arxiv.org/abs/1702.04792}{{\ttfamily 1702.04792}}].

\bibitem{Danielsson:2000mu}
U.~H. Danielsson, A.~Guijosa and M.~Kruczenski, \emph{{Newtonian gravitons and
  d-brane collective coordinates in wound string theory}},
  \href{https://doi.org/10.1088/1126-6708/2001/03/041}{\emph{JHEP} {\bfseries
  03} (2001) 041} [\href{https://arxiv.org/abs/hep-th/0012183}{{\ttfamily
  hep-th/0012183}}].

\bibitem{Buscher:1987sk}
T.~Buscher, \emph{{A Symmetry of the String Background Field Equations}},
  \href{https://doi.org/10.1016/0370-2693(87)90769-6}{\emph{Phys.\ Lett.\ B}
  {\bfseries 194} (1987) 59}.

\bibitem{Buscher:1987qj}
T.~Buscher, \emph{{Path Integral Derivation of Quantum Duality in Nonlinear
  Sigma Models}},
  \href{https://doi.org/10.1016/0370-2693(88)90602-8}{\emph{Phys.\ Lett.\ B}
  {\bfseries 201} (1988) 466}.

\bibitem{Bergshoeff:1994cb}
E.~Bergshoeff, R.~Kallosh and T.~Ortin, \emph{{Duality versus supersymmetry and
  compactification}},
  \href{https://doi.org/10.1103/PhysRevD.51.3009}{\emph{Phys. Rev. D}
  {\bfseries 51} (1995) 3009}
  [\href{https://arxiv.org/abs/hep-th/9410230}{{\ttfamily hep-th/9410230}}].

\bibitem{Bergshoeff:1996ui}
E.~Bergshoeff, M.~de~Roo, M.~B. Green, G.~Papadopoulos and P.~Townsend,
  \emph{{Duality of type II 7 branes and 8 branes}},
  \href{https://doi.org/10.1016/0550-3213(96)00171-X}{\emph{Nucl.\ Phys.\ B}
  {\bfseries 470} (1996) 113}
  [\href{https://arxiv.org/abs/hep-th/9601150}{{\ttfamily hep-th/9601150}}].

\bibitem{Bergshoeff:1995as}
E.~Bergshoeff, C.~M. Hull and T.~Ort{\'i}n, \emph{{Duality in the type II
  superstring effective action}},
  \href{https://doi.org/10.1016/0550-3213(95)00367-2}{\emph{Nucl.\ Phys.\ B}
  {\bfseries 451} (1995) 547}
  [\href{https://arxiv.org/abs/hep-th/9504081}{{\ttfamily hep-th/9504081}}].

\bibitem{Dai:1989ua}
J.~Dai, R.~Leigh and J.~Polchinski, \emph{{New Connections Between String
  Theories}}, \href{https://doi.org/10.1142/S0217732389002331}{\emph{Mod.\
  Phys.\ Lett.\ A} {\bfseries 4} (1989) 2073}.

\bibitem{Dine:1989vu}
M.~Dine, P.~Y. Huet and N.~Seiberg, \emph{{Large and Small Radius in String
  Theory}}, \href{https://doi.org/10.1016/0550-3213(89)90418-5}{\emph{Nucl.\
  Phys.\ B} {\bfseries 322} (1989) 301}.

\bibitem{Hull:1998vg}
C.~M. Hull, \emph{{Timelike T duality, de Sitter space, large N gauge theories
  and topological field theory}},
  \href{https://doi.org/10.1088/1126-6708/1998/07/021}{\emph{JHEP} {\bfseries
  07} (1998) 021} [\href{https://arxiv.org/abs/hep-th/9806146}{{\ttfamily
  hep-th/9806146}}].

\bibitem{Seiberg:1997ad}
N.~Seiberg, \emph{{Why is the matrix model correct?}},
  \href{https://doi.org/10.1103/PhysRevLett.79.3577}{\emph{Phys. Rev. Lett.}
  {\bfseries 79} (1997) 3577}
  [\href{https://arxiv.org/abs/hep-th/9710009}{{\ttfamily hep-th/9710009}}].

\bibitem{Brinkmann1}
H.~W. Brinkmann, \emph{{On Riemann Spaces Conformal to Euclidean Space}},
  \href{https://doi.org/10.1073/pnas.9.1.1}{\emph{Proceedings of the National
  Academy of Sciences} {\bfseries 9} (1923) 1}.

\bibitem{Alvarez-Gaume:1983ihn}
L.~Alvarez-Gaume and E.~Witten, \emph{{Gravitational Anomalies}},
  \href{https://doi.org/10.1016/0550-3213(84)90066-X}{\emph{Nucl. Phys. B}
  {\bfseries 234} (1984) 269}.

\bibitem{Bergshoeff:2020baa}
E.~Bergshoeff, A.~Chatzistavrakidis, J.~Lahnsteiner, L.~Romano and J.~Rosseel,
  \emph{{Non-Relativistic Supersymmetry on Curved Three-Manifolds}},
  \href{https://doi.org/10.1007/JHEP07(2020)175}{\emph{JHEP} {\bfseries 07}
  (2020) 175} [\href{https://arxiv.org/abs/2005.09001}{{\ttfamily
  2005.09001}}].

\bibitem{Gueven:1987ad}
R.~Gueven, \emph{{Plane Waves in Effective Field Theories of Superstrings}},
  \href{https://doi.org/10.1016/0370-2693(87)90254-1}{\emph{Phys. Lett. B}
  {\bfseries 191} (1987) 275}.

\bibitem{Bergshoeff:1992cw}
E.~A. Bergshoeff, R.~Kallosh and T.~Ortin, \emph{{Supersymmetric string
  waves}}, \href{https://doi.org/10.1103/PhysRevD.47.5444}{\emph{Phys. Rev. D}
  {\bfseries 47} (1993) 5444}
  [\href{https://arxiv.org/abs/hep-th/9212030}{{\ttfamily hep-th/9212030}}].

\bibitem{Bergshoeff:2017vjg}
E.~A. Bergshoeff, J.~Rosseel and P.~K. Townsend, \emph{{Gravity and the Spin-2
  Planar Schr\"odinger Equation}},
  \href{https://doi.org/10.1103/PhysRevLett.120.141601}{\emph{Phys. Rev. Lett.}
  {\bfseries 120} (2018) 141601}
  [\href{https://arxiv.org/abs/1712.10071}{{\ttfamily 1712.10071}}].

\bibitem{Callan:1991dj}
J.~Callan, Curtis~G., J.~A. Harvey and A.~Strominger, \emph{{World sheet
  approach to heterotic instantons and solitons}},
  \href{https://doi.org/10.1016/0550-3213(91)90074-8}{\emph{Nucl.\ Phys.\ B}
  {\bfseries 359} (1991) 611}.

\bibitem{Callan:1991at}
J.~Callan, Curtis~G., J.~A. Harvey and A.~Strominger, \emph{{Supersymmetric
  string solitons}},  \href{https://arxiv.org/abs/hep-th/9112030}{{\ttfamily
  hep-th/9112030}}.

\bibitem{Gross:1983hb}
D.~J. Gross and M.~J. Perry, \emph{{Magnetic Monopoles in Kaluza-Klein
  Theories}}, \href{https://doi.org/10.1016/0550-3213(83)90462-5}{\emph{Nucl.\
  Phys.\ B} {\bfseries 226} (1983) 29}.

\bibitem{Sorkin:1983ns}
R.~Sorkin, \emph{{Kaluza-Klein Monopole}},
  \href{https://doi.org/10.1103/PhysRevLett.51.87}{\emph{Phys.\ Rev.\ Lett.}
  {\bfseries 51} (1983) 87}.

\bibitem{Duff:1987bx}
M.~J. Duff, P.~S. Howe, T.~Inami and K.~S. Stelle, \emph{{Superstrings in D=10
  from Supermembranes in D=11}},
  \href{https://doi.org/10.1016/0370-2693(87)91323-2}{\emph{Phys. Lett. B}
  {\bfseries 191} (1987) 70}.

\bibitem{Kim:2007pc}
B.~S. Kim, \emph{{Non-relativistic superstring theories}},
  \href{https://doi.org/10.1103/PhysRevD.76.126013}{\emph{Phys. Rev. D}
  {\bfseries 76} (2007) 126013}
  [\href{https://arxiv.org/abs/0710.3203}{{\ttfamily 0710.3203}}].

\bibitem{Bergshoeff:1996tu}
E.~Bergshoeff and P.~K. Townsend, \emph{{Super D-branes}},
  \href{https://doi.org/10.1016/S0550-3213(97)00072-2}{\emph{Nucl. Phys. B}
  {\bfseries 490} (1997) 145}
  [\href{https://arxiv.org/abs/hep-th/9611173}{{\ttfamily hep-th/9611173}}].

\bibitem{Bigatti:1997jy}
D.~Bigatti and L.~Susskind, \emph{{Review of matrix theory}}, {\emph{NATO Sci.
  Ser. C} {\bfseries 520} (1999) 277}
  [\href{https://arxiv.org/abs/hep-th/9712072}{{\ttfamily hep-th/9712072}}].

\bibitem{Susskind:1997cw}
L.~Susskind, \emph{{Another conjecture about M(atrix) theory}},
  \href{https://arxiv.org/abs/hep-th/9704080}{{\ttfamily hep-th/9704080}}.

\bibitem{Hellerman:1997yu}
S.~Hellerman and J.~Polchinski, \emph{{Compactification in the lightlike
  limit}}, \href{https://doi.org/10.1103/PhysRevD.59.125002}{\emph{Phys. Rev.
  D} {\bfseries 59} (1999) 125002}
  [\href{https://arxiv.org/abs/hep-th/9711037}{{\ttfamily hep-th/9711037}}].

\bibitem{Bilal:1998vq}
A.~Bilal, \emph{{A Comment on compactification of M theory on an (almost)
  lightlike circle}},
  \href{https://doi.org/10.1016/S0550-3213(98)00203-X}{\emph{Nucl. Phys. B}
  {\bfseries 521} (1998) 202}
  [\href{https://arxiv.org/abs/hep-th/9801047}{{\ttfamily hep-th/9801047}}].

\bibitem{Lahnsteiner2022}
J.~Lahnsteiner, \emph{{Non-Lorentzian Supergravity and T-Duality}}, {\emph{to
  appear} (2022) }.

\bibitem{Hansen:2020pqs}
D.~Hansen, J.~Hartong and N.~A. Obers, \emph{{Non-Relativistic Gravity and its
  Coupling to Matter}},
  \href{https://doi.org/10.1007/JHEP06(2020)145}{\emph{JHEP} {\bfseries 06}
  (2020) 145} [\href{https://arxiv.org/abs/2001.10277}{{\ttfamily
  2001.10277}}].

\bibitem{Blair:2021waq}
C.~D.~A. Blair, D.~Gallegos and N.~Zinnato, \emph{{A non-relativistic limit of
  M-theory and 11-dimensional membrane Newton-Cartan geometry}},
  \href{https://doi.org/10.1007/JHEP10(2021)015}{\emph{JHEP} {\bfseries 21}
  (2021) 015} [\href{https://arxiv.org/abs/2104.07579}{{\ttfamily
  2104.07579}}].

\bibitem{Ebert:2021mfu}
S.~Ebert, H.-Y. Sun and Z.~Yan, \emph{{Dual D-Brane Actions in Nonrelativistic
  String Theory}},  \href{https://arxiv.org/abs/2112.09316}{{\ttfamily
  2112.09316}}.

\end{thebibliography}

\providecommand{\href}[2]{#2}\begingroup\raggedright\endgroup

\end{document}